\documentclass[11pt,a4paper]{article}
\usepackage[margin={1.3cm,1.3cm}]{geometry}
\usepackage{microtype}
\usepackage{graphicx}
\usepackage{authblk}
\usepackage[indent=10pt]{parskip}
\usepackage{enumerate}
\usepackage[normalem]{ulem}
\usepackage{hyperref}
\usepackage{upgreek}
\hypersetup{
    colorlinks=true,
    linkcolor=purple,
    filecolor=purple,      
    urlcolor=purple,
    citecolor=purple,
    pdftitle={Trust in society},
    pdfpagemode=FullScreen,
    }

\usepackage[font=footnotesize]{caption}
\usepackage[font=footnotesize]{subcaption}
\usepackage{booktabs} 
\usepackage{scalerel}

\usepackage{hyperref}

\usepackage{dsfont} 
\usepackage{algorithm}
\usepackage{mathrsfs}  
\usepackage{amsmath, bm}

\usepackage{amssymb}
\usepackage{mathtools}
\usepackage{amsthm}
\usepackage{xcolor}
\usepackage[thinc]{esdiff}

\newcommand{\rate}{\bm{\ell}}

\DeclareMathOperator{\E}{\mathbb E}

\DeclareMathOperator{\Cov}{\mathrm{Cov}}

\DeclareMathOperator{\diag}{diag}

\theoremstyle{plain}

\theoremstyle{definition}

\theoremstyle{remark}

\makeatother
\newcommand{\ee}{\mathbb{E}}
\newcommand{\vb}{\vspace{3.2mm}}

\title{Trust in society: A stochastic compartmental model}

\begin{document}





\author[1]{Benedikt V Meylahn\footnote{Corresponding author, mail: b.v.meylahn[at]uva.nl}}
\author[2]{Koen De Turck}
\author[1,3]{Michel Mandjes}

\affil[1]{Korteweg-de Vries Institute for Mathematics, University of Amsterdam, Amsterdam, The Netherlands}
\affil[2]{Department of Telecommunications and Information Processing, Ghent University, Ghent, Belgium}
\affil[3]{Mathematical Institute, Leiden University, P.O. Box 9512, 2300 RA Leiden, The Netherlands}

\date{}

\maketitle

\begin{abstract}
 This paper studies a novel stochastic compartmental model that describes the dynamics of trust in society. The population is split into three compartments representing levels of trust in society: trusters, skeptics and doubters. The focus lies on assessing the long-term dynamics, under `bounded confidence' (\textit{i.e.}, trusters and doubters do not communicate). We state and classify the stationary points of the system's mean behavior. We find that an increase in life-expectancy, and a greater population may increase the proportion of individuals who lose their trust completely. In addition, the relationship between the rate at which doubters convince skeptics to join their cause and the expected number of doubters is not monotonic --- it does not always help to be more convincing to ensure the survival of your group. We numerically illustrate the workings of our analysis. Because the study of stochastic compartmental models  for social dynamics is not common, we  in particular shed light on the limitations of deterministic compartmental models. 
 In our experiments we make use of fluid and diffusion approximation techniques as well as Gillespie simulation. 

\vb

\noindent
{\sc Acknowledgments.} This research was supported by the European Union’s Horizon 2020 research and innovation programme under the Marie Skłodowska-Curie grant agreement no. 945045, and by the NWO Gravitation project NETWORKS under grant no. 024.002.003. \includegraphics[height=1em]{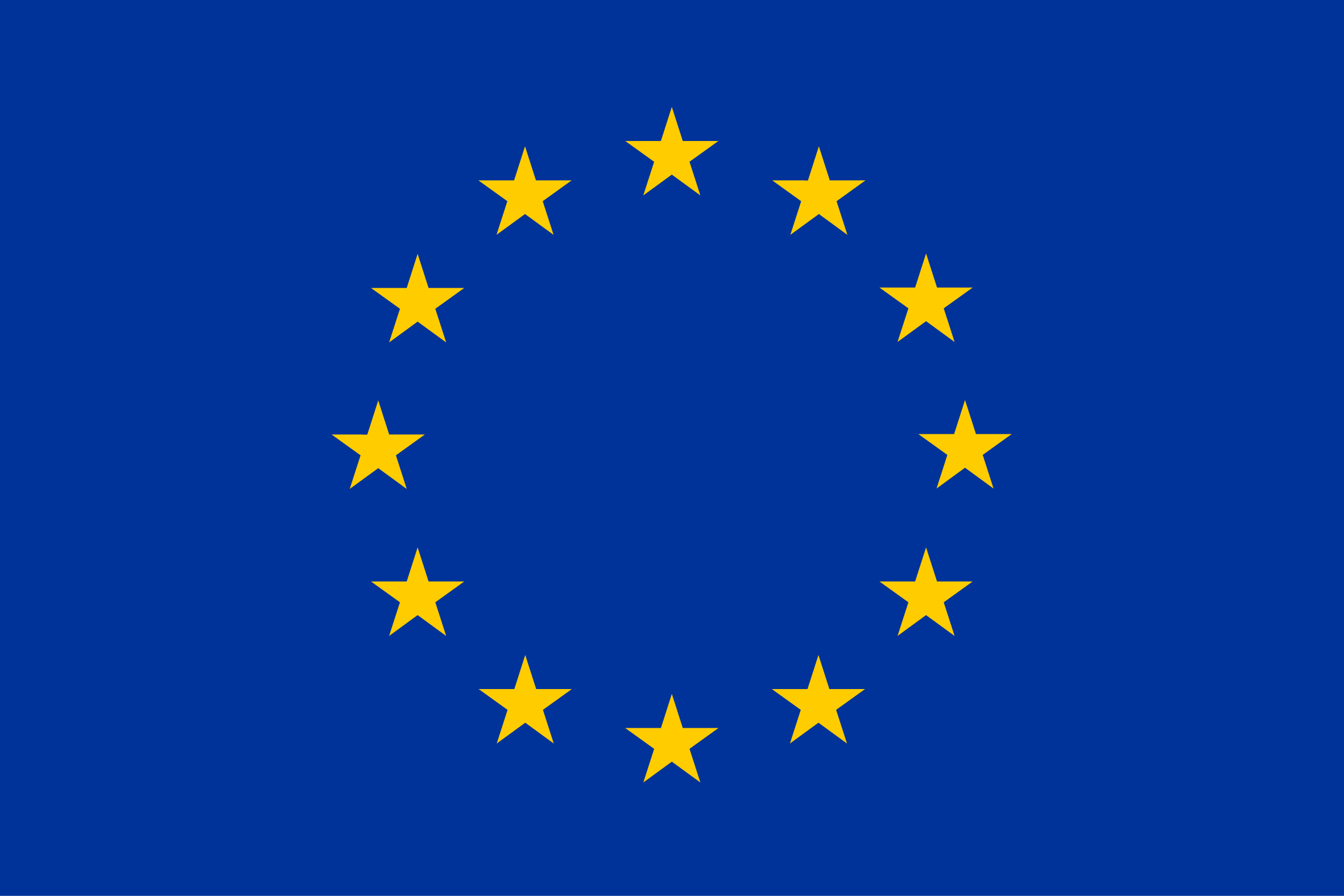}
Version: \today.
\end{abstract}
\section{Introduction}\label{sec:model}

Trust in the government is important for the healthy functioning of a society, yet a good amount of skepticism is important to ensure that this trust is not abused. This motivates our study of a new stochastic compartmental model for the dynamics of trust in society. A lack of trust in government and science has been linked to vaccine-hesitancy~\cite{Bajos2022,lazarus2022}. Meanwhile, the abundance of fake news during the COVID-19 pandemic~\cite{WHO2020} indicates that peer-to-peer influence may play a role in sowing distrust. In relation to such topics the populous may be rather divided and so peer-influence may in some cases give way to the concept of {\it bounded confidence}, \textit{i.e.}, individuals whose views are \textit{too} far apart no longer experience an assimilative force between them. Bounded confidence was introduced in the early 2000s~\cite{Deffuant2000, Hegselmann2002, Weisbuch2004}, and still receives attention~\cite{Altafini2018,Nugent2023,Nugent2024} (for reviews in the broader context of opinion dynamics models, we refer the reader to~\cite{Castellano2009,Chan2024,Proskurnikov2017,Proskurnikov2018,Noorazar2020}). 

In this paper we model the progression between individuals that trust fully (in the context of some institution \textit{e.g.}, a government), to those that have a healthy amount of skepticism, and finally those who have lost all trust. The model we present is a {\it stochastic compartmental model} that describes the three groups' random dynamics, incorporating the idea of bounded confidence. We note there is an abundance of works in the social sciences literature on compartmental models pertaining to the {\it deterministic} setting, while the stochastic setting is studied considerably less intensively. In this paper we argue that there are gains to be had working with stochastic compartmental models, as these provide insights beyond just the expected number of individuals in the individual compartments. This in particular opens up the option of studying probabilities of specific events (such as extinction, \textit{i.e.}, one of the compartments dying out) which do not happen in expectation yet with positive probability. 

At a more detailed level, our three-group stochastic compartmental model has the following dynamics. There is interaction between the trusters and the skeptics as well as between the skeptics and those who have lost trust. However, by bounded confidence, there is no interaction between those who have full trust and those who have lost it. By virtue of the idea that trust is much easier lost than gained we assume that the transition from skeptic to losing all trust is one-way, \textit{i.e.}, losing trust is permanent. We assume `deaths' (or departures) from all groups at the same rate, while `births' (or arrivals) occur only to the fully trusting compartment; birth and death rates are chosen such that the population remains roughly constant. Our choice that births lead to an increase of the trusting compartment corresponds to the situation in which the upbringing of the community assumes a high level of trust in society. This is an obvious simplification, but it is a reasonable starting point, reflecting that children learn to trust the government and society at school.

\subsection{Related literature}
We now present a non-exhaustive of the related literature, with a focus on papers from the social science literature (in the broad sense) on compartmental models for belief dynamics.

\subsubsection{General overview}
Models of epidemics \`a la Kermack and MacKendrick~\cite{Kermack1,Kermack2,Kermack3} describe the setting when individuals become infected with an idea and spread this to their neighbors. To model the spread of ideas in this way is not novel; in particular Daley and Kendall~\cite{Daley1965} model the stochastic spread of rumors through a workplace. A deterministic version of information spread has seen vast attention in the form of the Susceptible-Forwarding-Immune model and variations thereon~\cite{Yin2021,Yin2023,Magdaci2022} (see~\cite{Franceschi2022} for misinformation in particular). Bondesan \textit{et al.}\,\cite{Bondesan2024} take an integrated approach to jointly modelling vaccine-hesitancy and actual epidemic dynamics through a compartmental model. Departing from the idea of belief change, Kononovicius~\cite{Kononovicius2019} addresses the movement of agents with a fixed opinion between electorates. This provides an explanation to voting outcome change without assuming change of opinions held by individuals. 

The process of (de-)radicalization has been studied extensively using compartmental models~\cite{Chavez2003, Galam2016, Santoprete2017, Santoprete2018, Nathan2018, Santoprete2019}. A common thread in this branch of the literature is to model a general population and a core group. This core group consists of subclasses, \textit{viz.}\ susceptible, indoctrinated and radical. It is found that curbing the entry to the core group is crucial to curbing the growth of the radicalized sub-population~\cite{Chavez2003,Nathan2018}. Recently Wang~\cite{Wang2020} put forth a compartmental model in which a rational and an irrational idea compete to gain new followers. 

There is a line of literature~\cite{Dorso2017,daSilva2019,Ferrari2021,Kontorovsky2022} which addresses the intersection of beliefs and epidemics. In particular by modelling the effect of a belief model (usually done relying on agent-based models) on the spread of an epidemic model (which is of a compartmental nature). Dorso \textit{et al.}~\cite{Dorso2017} for instance, study the interplay of (anti-)vaccination behavior, public trust and the spread of measles.

\subsubsection{Directly related works}
There is an abundance of studies on compartmental models of social dynamics. In essence, these derive systems of ordinary differential equations, thus characterizing the {\it mean} behavior of the number of individuals in each sub-population as a function of time. As pointed out earlier, to gain a detailed understanding of the system's dynamic, one would like to incorporate its inherent random nature. In probabilistic terminology, the above-mentioned deterministic models effectively correspond to a \textit{fluid limit} (sometimes also referred to as the thermodynamic limit) of the stochastic model. Importantly, these fluid limits provide good approximations only when the population is sufficiently large. They are less appropriate in the regime of medium and smaller populations however, where the stochastic effects become more pronounced. This justifies working with a stochastic model, thus also incorporating the process' random behavior. Especially when the population size is far from the thermodynamic limit, the expected value of the dynamics tells a rather incomplete story. 

The models that mostly resemble (the deterministic version of) the dynamics we present are those by Wang~\cite{Wang2020} and McCartney and Glass~\cite{McCartney2015}. As mentioned, Wang~\cite{Wang2020} studies the dynamics by which individuals switch from not thinking much on a topic to either holding a rational or an irrational view on the topic. He incorporates bounded confidence by setting the influence between rational and irrational view holders to zero. The transition from holding an irrational view to holding the rational one thus takes place by first having a moment of self-realization which induces doubt in the irrationalist. The doubters may subsequently be influenced by individuals holding the rational view. McCartney and Glass~\cite{McCartney2015} present a model in which individuals switch from being committed to a church, to non-committed and finally non-affiliated. They do not explicitly make use of bounded confidence, though upon fitting their model to data, they empirically find that the influence between the committed church goers and the non-affiliated individuals is close to zero. Though not mentioned explicitly, it is conceivable that the decisions individuals take to switch from one group to another in their model may also be heavily influenced by trust. 

Recently Nugent \textit{et al.}~\cite{Nugent2024} illustrated, in the context of opinion dynamics with bounded confidence, how increasing the interaction rate while decreasing the impact of an interaction in agent-based models results in convergence of the random model dynamics to solutions of differential equations. Their work is related in the sense that they too explicitly discuss the connection between the deterministic differential equation approach and their stochastic counterpart. While both their work and ours is concerned with the evolution of beliefs in a population, their results are not directly applicable to the model we consider because of a fundamental difference: the agents in our model fall into discrete compartments which describe their belief level, while their model considers the belief of each individual agent separately. 

\subsection{Contributions}
A main contribution of our work concerns the new stochastic compartmental model for the dynamics by which individuals in a society transition between, high, moderate and no trust. In formulating this model we have taken care to incorporate the reasonable assumption of bounded confidence, as is typically used in the literature. The model we present is conceptually simple yet behaviorally rich. We identify and classify the three stationary points of the fluid limit of this model, resulting in a phase plane in the parameter space indicating which (unique) stationary point is stable.

By using our model, we gain insights about the dynamics of trust in society. We observe a surprising relationship between the rate at which those who have lost all trust convince others of their view on one hand, and the expected number of individuals who have lost all trust on the other hand. In particular this relationship is not monotonic; there is a point at which increasing the rate whereby doubters convince skeptics decreases the expected number of doubters. The model also predicts that a longer life-span and a larger population both increase the expected proportion of individuals who have lost trust. Because we study a stochastic model, we are also able to analyze the impact of the model  parameters on the extinction probabilities of the individual groups.

Finally, our paper illustrates the application of computational methods used in the study of stochastic social compartmental models. As mentioned, substantial understanding can be gained by performing a stochastic analysis, rather than relying on the differential equations that underlie the model's deterministic counterpart, in particular if the groups that are being modeled are not large. The method of diffusion approximation readily provides us with a quantification of relevant probabilities, such as the probability of reaching an unusually high value of individuals in a certain compartment. This is useful for instance when modeling extremism dynamics (a topic in which deterministic compartmental models have been readily applied), where having an extremist group of a critical size can have important implications for the activities of that group. In instances that the diffusion approximation runs into conceptual issues (as it cannot properly handle extinction of a critical compartment), we study the stochastic dynamics by means of a stochastic simulation algorithm.

\subsection{Organization of paper}
In this introductory section we have motivated the study of a stochastic compartmental model for the dynamics by which individuals lose and gain trust in institutions. In \S\ref{sec:gen_model} we introduce the notation used. We present the methods of fluid and diffusion approximation in \S\ref{sec:methods}, as well as Doob-Gillespie simulation. In \S\ref{sec:analysis} we present the analysis of the model and conclude in \S\ref{sec:discussion} with a discussion of the results, as well as an account of the limitations of the model and possible future work.

\section{General model notation}\label{sec:gen_model}
Because of the broad applicability of the techniques we use to study our compartmental model, we separate the presentation of the main example to which we apply those techniques and the presentation of the techniques themselves. 

\subsection{General setup}
We study a subclass of the population processes considered in Kurtz~\cite{Kurtz1981}, and as such we have a substantial overlap in notation. Let, for some $d\in{\mathbb N}$, $\bm{X}(t)=[X_1(t), X_2(t),\ldots,X_d(t)]^\top$ denote a vector-valued stochastic population process in continuous time $t\in [0,\infty)$. The value $X_i(t)\in {\mathbb N}_0$ denotes the number of individuals in compartment $i\in\{1,\ldots,d\}$ at time $t$, where $d$ is the number of compartments in the model. In the following, we will make use of the notation ${\bm e}_i$ to denote the vector with a one on position $i$ and zeroes elsewhere. The dynamics in the model are due to events which occur at Poisson times. There are three different types of events:
\begin{enumerate}
    \item \textit{Arrival}: The new individual arrives at a compartment, thus increasing the latter's size by 1. If this concerns compartment $i\in\{1,\ldots,d\}$, then the change is in direction ${\bm\ell}=\bm{e}_i$.
    \item \textit{Exit}: An individual exits the system, thus decreasing the size of one compartment by 1. If this concerns compartment $i\in\{1,\ldots,d\}$, then the change is in direction ${\bm\ell}=-\bm{e}_i$.
     \item \textit{Jump}: An individual moves from one compartment to another, thus decreasing and increasing the size of the two compartments in question by 1 respectively. If this concerns a transition from compartment $i\in\{1,\ldots,d\}$ to compartment $j\in\{1,\ldots,d\}\setminus\{i\}$, then the change is in direction ${\bm\ell}=-\bm{e}_i+\bm{e}_j$.
\end{enumerate}
Observe that, evidently, in the cases of an exit or jump from compartment $i$, the number of individuals in compartment $i$ should be strictly positive. Hence, the set of possible directions is
\[{\mathsf L}:=\bigcup_{i=1}^d\{\bm{e}_i\}\,\cup\, \bigcup_{i=1}^d\bigcup_{j\not= i}\{-\bm{e}_i+\bm{e}_j\} \,\cup\, \bigcup_{i=1}^d\{-\bm{e}_i\}.\]
We conclude that in this general setup there are $d(d+1)$ possible transition directions. In the sequel, $q_{\bm{\ell}}(\bm{X}(t))$ denotes the transition rate of the underlying continuous time Markov chain if at time $t$ the system state is $\bm{X}(t)$.

Transition rates may be either independent of, or proportional to the size of the compartments that they involve. As our compartmental model does not require the highest level of generality, we focus on the following specific case:
\begin{itemize}
    \item[$\circ$] The arrival rates are independent of the state of the process. An arrival to compartment $i\in\{1,\ldots,d\}$ occurs at rate $\lambda_i$. Hence the transition rate $q_{\bm{\ell}}$ for ${\bm\ell}={\bm e}_i$ reads, with ${\bm X}(t)$ the system state at time $t$,
    \begin{equation}
        q_{\bm{e}_i}({\bm X}(t))= \lambda_i,
    \end{equation}
    for $\lambda_i\geqslant 0$.
     \item[$\circ$]  The exit rates are dependent only on the size of the compartment from which the individual is exiting. An exit from compartment $i\in\{1,\ldots,d\}$ occurs, per individual that is present, at rate $\mu_i$. Hence the transition rate $q_{\bm{\ell}}$ for ${\bm\ell}=-{\bm e}_i$ reads, with ${\bm X}(t)$ the system state at time $t$,
    \begin{equation}
        q_{-\bm{e}_i}({\bm X}(t))= \mu_iX_i(t),
    \end{equation} 
    for $\mu_i\geqslant 0.$
    Observe that the individuals exit compartment $i$ independently of each other. If $X_i(t)=0$ this rate is automatically $0$, as desired.
     \item[$\circ$]  The jump rates are dependent on the size of the compartment from which the individual is exiting as well as the size of the compartment it is joining. A jump from compartment $i\in\{1,\ldots,d\}$ to compartment $j\in\{1,\ldots,d\}\setminus\{i\}$ occurs, per pair of individuals that is present in each of the two compartments, at rate $\nu_{ij}$. Hence the transition rate $q_{\bm{\ell}}$ for ${\bm\ell}=-{\bm e}_i+{\bm e}_j$ reads, with ${\bm X}(t)$ the system state at time $t$,
    \begin{equation}
        q_{-\bm{e}_i+{\bm e}_j}({\bm X}(t))= \nu_{ij}X_i(t)X_j(t),
    \end{equation} 
    for $\nu_{ij}\geqslant 0.$ If $X_i(t)=0$ or $X_j(t)=0$, then this rate is $0$.
\end{itemize}

We have now set up our framework, in that for each event $\bm{\ell}\in \mathsf{L}$ we can characterize the probability that it takes place in an interval of length $\Delta$. Indeed, for $\Delta$ small,
\begin{equation}
    \mathbb{P}\left(\bm{X}(t+\Delta) = \bm{X}(t) + \bm{\ell} \mid \bm{X}(s), s\leqslant t\right) = q_{\bm{\ell}}(\bm{X}(t))\,\Delta +o(\Delta).
\end{equation}
The generator ${\mathscr A}$ of the underlying Markov process is therefore given via 
\begin{equation}
    {\mathscr A}\,\bm{F}(\bm{x}) = \sum_{\bm{\ell}\in\mathsf{L}}q_{\bm{\ell}}(\bm{x})(\bm{F}(\bm{x}+\bm{\ell})-\bm{F}(\bm{x})).\label{eq:gen}
\end{equation}

\subsection{Compartmental model}\label{sec:mainmodel}
\begin{figure}
    \centering
    \includegraphics{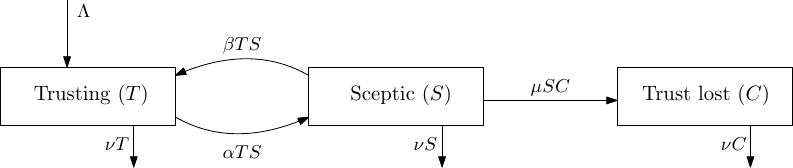}
    \caption{Stochastic flow diagram of the compartmental trust model.}
    \label{fig:flow}
\end{figure}

We now use the general model described in the previous subsection to introduce the compartment model of our interest.
We study the stochastic population process given by the vector $\bm{X}(t) = [T(t), S(t), C(t)]^\top \in \mathbb{N}_{0}^3$ over continuous time $t\geqslant 0$. We have divided the population into those who
\begin{itemize}
    \item[$\circ$] trust the institution, referred to as $T(t)$,
    \item[$\circ$] are skeptical about the institution, referred to as $S(t)$, and
    \item[$\circ$] have lost trust in the institution, referred to as $C(t)$.
\end{itemize}
We assume that by random meetings trusters and skeptics may convince one another of their views. Similarly by a meeting between a skeptic and a doubter the skeptic may lose all trust and join the doubters. Thus transitions between times $t$ and $t+\Delta$ occur with probabilities proportional to $\Delta$, as $\Delta\downarrow 0$. Indeed,  for $\Delta$ small, if the current state is $(T,S,C)$, we have
\begin{alignat}{4}
    (T,S,C)&\to (T-1,S+1,C), \quad &&\text{w.p. }\alpha\, TS\,\Delta  +o(\Delta)\quad &&\text{(sowing seeds of doubt),}\label{eq:doubt}\\
   (T,S,C)&\to (T+1,S-1,C),   \quad&&\text{w.p. }\beta\, TS\,\Delta  + o(\Delta)\quad&&\text{(reassurance),}\label{eq:reassure}\\
    (T,S,C)&\to (T,S-1,C+1),  \quad &&\text{w.p. }\mu\,SC\,\Delta  +o(\Delta)\quad && \text{(trust lost)}\label{eq:loss}.
\end{alignat}
Furthermore we let there be births only at the trusters compartment. The justification behind this choice lies in the fact that most individuals born to a population are subject to some kind of schooling where they would in general be exposed to the status quo which we assume to be trust in the institutions. The Poisson process by which new individuals are added to the population of trusters has a constant rate $\Lambda>0$. As motivated in the introduction, we have chosen to not have any external arrivals to the classes $S$ and $C$. We also allow individuals from all compartments to leave the population, which happens at rate $\nu X$ for each compartment $X\in \{T,S,C\}$. The model is visualized in the stochastic flow diagram of  Figure~\ref{fig:flow}, revealing that in our specific instance we have that $|{\mathsf L}|=7$, and
\begin{equation}
    {\mathsf L} = \left\{\left(\begin{array}{r}1\\0\\0\end{array}\right),
    \left(\begin{array}{r}-1\\0\\0\end{array}\right),
    \left(\begin{array}{r}0\\-1\\0\end{array}\right),
    \left(\begin{array}{r}0\\0\\-1\end{array}\right),
    \left(\begin{array}{r}-1\\1\\0\end{array}\right),
    \left(\begin{array}{r}1\\-1\\0\end{array}\right),
    \left(\begin{array}{r}0\\-1\\1\end{array}\right)\right\}.
\end{equation}

Note that we do not include a transition from having lost trust to either of the other compartments in the current model. This is in the vein of trust being easily lost and hard to regain~\cite{Lewicki1996,Lewicki2000}. This decision is motivated by empirical evidence, \textit{e.g.}\ in the findings of McCartney and Glass~\cite{McCartney2015}. When fitting their models of church affiliation to census data of Northern Ireland, they found that the transition rate of those who had lost their faith back to the church was either zero or close to zero.

Recently empirical research has shifted to investigate the effectiveness of strategies aimed at repairing trust once it has been lost (see for example~\cite{Solinger2016,Guo2018,Zhou2024}). With active efforts it seems that trust may be repaired. Thus, inclusion of institutional efforts at rebuilding trust in the model is a promising direction of future work.


\section{Methodology}\label{sec:methods}
In this section we discuss methodologies to assess the stochastic model defined in the previous section. Models of this kind admit both a {\it fluid} and a {\it diffusion} approximation. The fluid approximation can be considered as a {\it law of large numbers} at the process level: informally, under a scaling in which the number of individuals in the system grows large, the evolution of each of the components tends to a deterministic function, which can often be characterized as the solution to a set of ordinary differential equations. The diffusion approximation, on the other hand, concerns the fluctuations around the fluid limit and can be seen as a {\it central limit theorem} at the process level. In cases where these approximations fail the {\it Doob-Gillespie stochastic simulation algorithm} provides a means to sample statistically correct trajectories of the process, and can therefore be used to estimate any relevant quantity. The failure of the diffusion approximation relates to transitions out of absorbing states and is elaborated on in \S\ref{sec:probSkep_die}. In short, by the diffusion approximation does not account for the fact that if at some point the $S$ (or the $C$) compartment is empty, that this stays true indefinitely. In the rest of this section we provide brief accounts of these three methods, which we will intensively apply in Section \ref{sec:analysis} to study our specific compartmental model.

\subsection{Fluid approximation}
In the fluid limit we scale the process $\bm{X}$ by a factor $N$, to be thought if as being on the order of the size of the total population. The process of interest is the {\it proportion} (or density) \begin{equation}\bm{X}_N(t):=\frac{\bm{X}(t)}{N}.
\end{equation} 
In order to keep transitions occurring at an appropriate pace we also have to scale the transition rates, in that all of them are multiplied by a factor $N$; we refer to the rescaled rates by $\beta_{\bm \ell}(\cdot)$. 
Then observe that this means that, as $\Delta\downarrow 0$,
\begin{align}\notag
    \frac{{\mathbb E}[T_N(t+\Delta)-T_N(t)\,|\,{\bm X}_N(t)]}{\Delta}&= \Lambda - \nu \,T_N(t) + \beta\, T_N(t)S_N(t) -\alpha\, T_N(t)S_N(t) + o(1),\\
    \frac{{\mathbb E}[S_N(t+\Delta)-S_N(t)\,|\,{\bm X}_N(t)]}{\Delta}&=  - \nu \,S_N(t) - \beta\, T_N(t)S_N(t) +\alpha\, T_N(t)S_N(t) - \mu\, S_N(t)C_N(t)+ o(1),\label{eqs}\\
    \frac{{\mathbb E}[C_N(t+\Delta)-C_N(t)\,|\,{\bm X}_N(t)]}{\Delta}&= - \nu\, C_N(t) + \mu\, S_N(t)C_N(t) + o(1).\notag
\end{align}
Sending $\Delta\downarrow 0$ in these equations, the above display suggests that ${\bm X}_N(t)$ converges to the solution of a system of ordinary differential equations. In the remainder of this subsection we argue that this is indeed the case.  

Following closely the procedure leading up to  Kurtz~\cite[Theorem~8.1]{Kurtz1981}, and the proof thereof, we define
\begin{equation}
    \bm{F}(\bm{x}):= \sum_{\bm{\ell}\in \mathsf{L}}\bm{\ell}\,\beta_{\bm{\ell}}(\bm{x}),
\end{equation}
which is a column vector of length $d$ with the rates corresponding to the transitions in the scaled model were it in state $x\in \mathbb{R}_{\geqslant 0}^d.$ We define the {\it fluid path} $\bm{x}(\cdot)$ as the solution to the following $d$ coupled ordinary differential equations:
\begin{equation}
    \diff{\bm{x}(t)}{t} = \bm{F}(\bm{x}(t)), \quad \bm{x}(0) = \bm{x}_0;\label{eq:fluid_de}
\end{equation}
observe the similarity with the system \eqref{eqs}. 
By the results in \cite{Kurtz1981}, the process $\bm{X}_N(\cdot)$ converges to the fluid path $\bm{x}(\cdot)$ as $N\to\infty$. Formally, if the conditions
\begin{enumerate}[\hspace{1em}{F}.1]
   \item $\lim_{N\to \infty}\bm{X}_N(0)=\bm{x}_0$,
    \item $\sum_{\bm{\ell}\in\mathsf{L}}\sup_{\bm{x}\in\bm{K}}\beta_{\bm{\ell}}(\bm{x}) <\infty$, for any compact subset $\bm{K}$ of $\mathbb{R}^d$, and
        \item there exists $M_K>0$ such that $ |\bm{F}(\bm{x})-\bm{F}(\bm{y})|\leqslant M_K\,|\bm{x}-\bm{y}|$, for all $\bm{x},\bm{y}\in \bm{K}$, for any compact subset $\bm{K}$ of $\mathbb{R}^d$ 
\end{enumerate}
apply, then we have that, for all $t\geqslant 0$,
\begin{equation}
    \lim_{N\to\infty} \sup_{s\leqslant t}|\bm{X}_N(s)-\bm{x}(s)| = 0.
\end{equation}
We provide the verification of conditions F.1--F.3 in Appendix~\ref{ap:verificationF}. 

The resulting system of ordinary differential equations (\ref{eq:fluid_de}) corresponds to the system that one would obtain when working with {\it deterministic} flows of the same rates as in our stochastic compartment model. The added value of our  stochastic model is that more refined objects, \textit{i.e.}, quantities that are more detailed than just the expectation, are also tractable. For this we turn to the diffusion approximation, shedding light on typical deviations from the mean, and later to stochastic simulation.
\subsection{Diffusion approximation}
We proceed by analyzing the typical fluctuations around the fluid path. 
To this end, we define by $\hat{\bm{X}}_N(t):= \sqrt{N}(\bm{X}_N(t)-\bm{x}(t))$, for $t\geqslant 0$, the scaled difference between our scaled process and the associated fluid path. Furthermore we define the {\it Jacobian matrix}
\begin{equation}
    \nabla \bm{F}(\bm{x}) := 
    \begin{bmatrix} 
    \diffp{\bm{F}(\bm{x})_1}{{x_1}} & \cdots & \diffp{\bm{F}(\bm{x})_1}{{x_d}} \\ 
    \vdots & \ddots & \vdots \\ 
    \diffp{\bm{F}(\bm{x})_d}{{x_1}}& \cdots & \diffp{\bm{F}(\bm{x})_d}{{x_d}}
    \end{bmatrix}
\end{equation}
of dimension $d\times d$. 
In addition there is the \textit{transition matrix} $H$, of dimension $d\times |\mathsf{L}|$, containing the possible transitions as columns, being the direct sum of all the transition vectors
\begin{equation}
    H := \bigoplus_{\bm{\ell}\in\mathsf{L}}{\bm \ell}.
\end{equation}
Finally we define the square \textit{rates matrix}, $\Sigma$, of dimension $|\mathsf{L}|\times |\mathsf{L}|$,
\begin{equation}
    \Sigma(\bm{x}):=\diag\left(\sqrt{\beta_1(\bm{x})}, \ldots ,\sqrt{\beta_{|\mathsf{L}|}(\bm{x})}\right).
\end{equation}
The {\it diffusion limit} $\hat{\bm{X}}(\cdot)$ is a $d$-dimensional process that obeys the stochastic differential equation, with ${\bm x}(\cdot)$ as before denoting the fluid path, 
\begin{equation}
 \text{d}\hat{\bm{X}}(t) = \nabla \bm{F}(\bm{x}(t))\,\hat{\bm{X}}(t)\,\text{d}t + H\Sigma(\bm{x}(t))\,\text{d}{\bm W}(t),\label{eq:dif_sde}
\end{equation}
where ${\bm W}(\cdot)$ is an $|{\mathsf L}|$-dimensional process whose entries are independent standard Brownian motions. 
Then we can appeal to the diffusion limit result of Kurtz~\cite[Theorem~8.2]{Kurtz1981} to conclude that
\begin{equation}
    \hat{\bm{X}}_N(\cdot) \stackrel{\rm d}{\to}\hat{\bm{X}}(\cdot), 
\end{equation}
with `$\stackrel{\rm d}{\to}$' denoting distributional convergence,
if
\begin{enumerate}[\hspace{1em}{D}.1]
    \item $\lim_{N\to\infty}\sqrt{N}|\bm{X}_N(0)-{\bm x}_0|=0$ almost surely,
    \item $\sum_{\bm{\ell}\in\mathsf{L}}|\bm{\ell}|^2\sup_{\bm{x}\in\bm{K}}\beta_{\bm{\ell}}(\bm{x}) <\infty$, for all compact $\bm{K}\subset\mathbb{R}^d$ and
    \item $\nabla \bm{F}(\bm{x})$ is a bounded, continuous function of $\bm{x}$.
\end{enumerate}
We provide the verification of conditions D.1--D.3  in the context of our $3$-dimensional compartmental model in Appendix~\ref{ap:verificationD}.  

With the help of formulas derived in Karatzas and Shreve~\cite[\S5.6]{Karatzas1988}, we can deduce the covariance matrix of an SDE with the form displayed in (\ref{eq:dif_sde}). Defining, for any $s,t\geqslant 0$, by $\bm\Gamma(t,s):=\Cov[\hat{\bm{X}}(t),\hat{\bm{X}}(s)]$ the $d\times d$ covariance matrix, we have
\begin{equation}
    \bm\Gamma(t,s)= {\bm \Phi}(t)\left[\bm\Gamma(0,0)+\int_{0}^{\min\{t,s\}}{\bm \Phi}^{-1}(u)H\Sigma(\bm{x}(u)) \times \left(\bm\Phi^{-1}(u)H\Sigma(\bm{x}(u))\right)^\top\text{d}u\right]{\bm \Phi}(s)^\top,
\end{equation}
where the $d$-dimensional function ${\bm \Phi}(\cdot)$ solves the system of coupled ordinary differential equations, with $t\geqslant 0$,
\begin{equation}
    \diff{\bm \Phi(t)}{t} = \nabla \bm{F}(\bm{x}(t))\,{\bm \Phi}(t), \quad {\bm \Phi}(0) = I_d,
\end{equation}
and $I_d$ denotes the $d$-dimensional identity matrix. 
A simpler solution, which is particularly useful when one wishes only to know covariances between compartments at the \textit{same} point in time, results from the system of differential equations
\begin{equation}
    \diff{{\bm V}(t)}{t} = \nabla \bm{F}(\bm{x}(t)){\bm V}(t) + {\bm V}(t) \left(\nabla \bm{F}(\bm{x}(t))\right)^\top + H\Sigma(\bm{x}(t))\times \left(H\Sigma(\bm{x}(t))\right)^\top,
\end{equation}
where $\bm V(t) :=\bm\Gamma(t,t)$. 

Upon combining the above, the grand approximation for the scaled process is given by, for $t\geqslant 0$, \begin{equation}
    \bm{X}_N(t) \approx \bm{x}(t)+\frac{\hat{\bm{X}}(t)}{\sqrt{N}}.
\end{equation}
Because the first term on the right-hand side is deterministic and the second term has mean 0, 
the mean of this process follows the fluid path, while the covariance between $\bm{X}_N(t)$ and $\bm{X}_N(s)$ is approximated by $\Gamma(t,s)/\sqrt{N}$.


\subsection{Gillespie simulation}\label{sec:sim}
By applying the diffusion approximation we can produce a proxy of the probability of our compartment process being in some set (for instance: the set of all paths such that $S(t)$ remains below a certain threshold $\bar S$ for all $s\leqslant t$, for a given $t>0$). The diffusion approach has its limitations, though, the most prominent one being related to the event of extinction. For instance, if $S(t)$ becomes $0$ for some $t$, it cannot become positive again, and also $C(t)$ will eventually die out; evidently, this effect cannot be incorporated in the diffusion scaling. To analyze this type of events, stochastic simulation is a more appropriate tool.

The {\it Doob-Gillespie algorithm}, also known as the {\it stochastic simulation algorithm}~\cite{Gillespie1977}, provides a convenient procedure for sampling statistically correct trajectories of a system in which the rates at which events occur are known. It is particularly popular in the simulation of chemical reaction networks.

The algorithm depends on the following well-known result in the study of Poisson processes. Consider $k$ events occurring independently at exponentially distributed times $T_1,\ldots, T_k$, respectively, with $q_1,\ldots,q_k$ denoting the corresponding rates. Then it holds that the earliest of these events, \textit{i.e.},
\begin{equation}
    T:=\min\{T_i: i=1,\ldots,k\},
\end{equation}
occurs at a time which is exponentially distributed with rate equal to the sum of the rates $q=\sum_{i=1}^k q_i$. The probability that event $i$ is this earliest event is $q_i/q$ for all $i\in\{1,\ldots, k\}$, independently of the value of $T$. This is a fundamental result in the study of Poisson processes; for a formal statement and more background, see \textit{e.g.}\ Norris~\cite[Theorem 2.3.3]{Norris1997}).

This result can be used to simulate our compartment process by repeatedly following the steps:
\begin{enumerate}
    \item Sum all rates of the stochastic process based on the state of the system $q=\sum_{\bm{\ell}}\beta_{\bm{\ell}}(\bm{x}(t))$.
    \item Draw a time increment $\uptau$ exponentially distributed at rate $q$.
    \item Draw an event $\bm{\ell}$ proportional to $\beta_{\bm{\ell}}/q$.
    \item Update the system according to the transition of event $\bm{\ell}$: $\bm{x}\leftarrow \bm{x}(t) + {\bm\ell}$, and update the time $t\leftarrow t+\uptau.$
\end{enumerate}
This simulation procedure simulates a number of transitions rather than a number of time steps. However, in order to aggregate simulation runs, the state of the system at specific points in simulated time need to be stored for comparison. To do this
we run each iteration for $200\times N$ transitions. During the iteration we capture the state of the system at time $t\in\{1,6,11,16,\ldots,296\}$ as well as the final state. We run 500 iterations of each parameter setting we mention.
\subsection{Alternative methods}
We conclude this section by  briefly discussing alternative methods which are potentially useful in the numerical evaluation of compartmental model, although we will not work with them in studying our model.

The method of {\it moment closure approximations} is commonly applied in the study of chemical reaction networks~\cite{Schnoerr2015,Schnoerr2017}, and explored in greater detail by Kuehn~\cite{Kuehn2016}. The central idea in a moment closure approximation is to transform an infinite system of differential equations into a finite system by assuming that the process vector has a specified distributional form.

It is also possible to use the {\it diffusion approximation in combination with simulation}. The idea is that one approximates the system at hand via a diffusion limit (in the way we have demonstrated above), but then to use simulation tools to numerically assess it. In particular, when analyzing an SDE it is common to simulate trajectories thereof by means of the Euler-Maruyama method; extensive introductions to this method are available in {e.g.}~\cite{Kloeden1992,Asmussen2007}. This may offer advantages in terms of efficiency over the Doob-Gillespie simulation because instead of simulating individual changes of the state in the Markov process, one chooses a grid or discretization and simulates steps of the chosen size. However, care may be necessary to avoid facing issues similar to those in the direct diffusion approximation; as pointed out earlier, diffusion-based tools have the intrinsic problem of allowing transitioning out of an absorbing state.

\section{Model analysis}\label{sec:analysis}
In this section we analyze the model presented in \S\ref{sec:mainmodel} by means of fluid and diffusion approximation as well as stochastic simulation. 
In particular we are interested in the interplay between the model parameters and the probability of long-term survival of the doubter group. 

Throughout our study we consider the regime in which the total population remains essentially constant, which we enforce by taking 
$\nu N=\nu(T+S+C)=\Lambda$. For simplicity we set $N=1$, so that $\Lambda = \nu$. This results in the fluid path being characterized as the solution of the following system of ordinary differential equations: in our setting, \eqref{eq:fluid_de} becomes
\begin{align}
    \dot{T}_f &=\Lambda -\nu T_f -\alpha T_f S_f +\beta T_f S_f\\
    \dot{S}_f &=\alpha S_f T_f -\beta T_f S_f- \nu S_f -\mu S_f C_f\\
    \dot{C}_f&=\mu S_f C_f -\nu C_f.
\end{align}
We introduce the subscript $f$ to indicate that we have taken the fluid limit. We can somewhat simplify this system by defining $\gamma:= \alpha-\beta$. Upon inspecting the nature of these differential equations, we note that if $\gamma<0$, then $S_f$ dies out: the rate at which individuals leave from this compartment exceeds the rate at which they enter. If $S_f$ dies out, so does $C_f$ eventually, because the $S_f$ compartment is the only one that feeds $C_f$. Thus we proceed the analysis of the model assuming that $\gamma>0$. Additionally we note that if $\nu>\gamma$, then $S_f$ and $C_f$ both die out because $\gamma S_f T_f-\nu S_f<0$. Similarly we note that if $\mu<\nu$, then $C_f$ dies out asymptotically because then $\mu S_f C_f-\nu C_f <0$ which implies that there is a drift toward $C_f=0$.

We can make another simplification because we have set the arrival rate equal to the (total) exit rate. Thus, in this fluid limit regime, we may replace $C_f$ by $1-S_f-T_f$, so as to obtain the following simplified system of differential equations:
\begin{align}
    \dot{T}_f &=\nu -\nu T_f -\gamma T_f S_f \\
    \dot{S}_f &=(\gamma +\mu)S_f T_f - (\nu+\mu)S_f +\mu S_f^2.
\end{align}

\subsection{Stationary point analysis}\label{sec:steady}
Setting $\dot{\bm{X}}$ to zero, and solving the resulting three equations simultaneously, we conclude there are three stationary points: \begin{equation}
    \bm{X}_1=(1,0,0),\:\:\:\bm{X}_2=\left(\frac{\nu}{\gamma},1-\frac{\nu}{\gamma},0\right),\:\:\:\bm{X}_3=\left(\frac{\mu}{\gamma+\mu},\frac{\nu}{\mu},\frac{\gamma }{\gamma+\mu}-\frac{\nu}{\mu} \right).
\end{equation}
With
\begin{equation}
    \kappa := \frac{\gamma^{3/2}\sqrt{\nu} +\gamma\nu}{2(\gamma-\nu)},
\end{equation}and $\nu$, these can be classified as follows:
\begin{itemize}
    \item[$\circ$] $\bm{X}_1$ is saddle point when $\nu<\gamma $ and a stable node when $\gamma >\nu$.
    \item[$\circ$] $\bm{X}_2$ is a saddle point when $\gamma<\nu$ or when $\gamma >\frac{\mu \nu}{\mu-\nu}$. It is a stable node or star in between the two.
    \item[$\circ$] $\bm{X}_3$ is a saddle point when $\mu <\frac{\gamma \nu}{\gamma-\nu}$, transitions into a stable node when $\mu >\frac{\gamma \nu}{\gamma-\nu}$ and transitions again into a stable spiral when $\mu>\kappa$.
\end{itemize}
    To show this we inspect the stability of these points in the linearized system. We provide more details of this in Appendix~\ref{app:ODE_stab} which closely follows the procedure outlined in~\cite[Chapter 4]{Strogatz2024}. 

\begin{figure}[htb]
    \centering
    \begin{subfigure}{0.6\textwidth}
    \centering
        \includegraphics{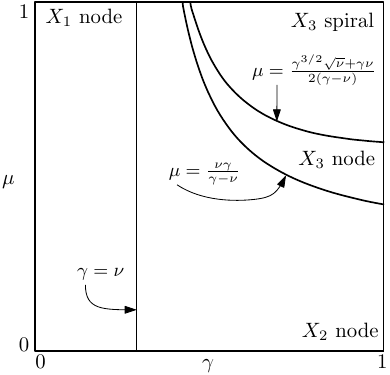}
        \caption{Phase plane} \label{fig:Stability} 
\end{subfigure}%
        \begin{subfigure}{0.333\textwidth}
    \centering
        \includegraphics[width = \textwidth]{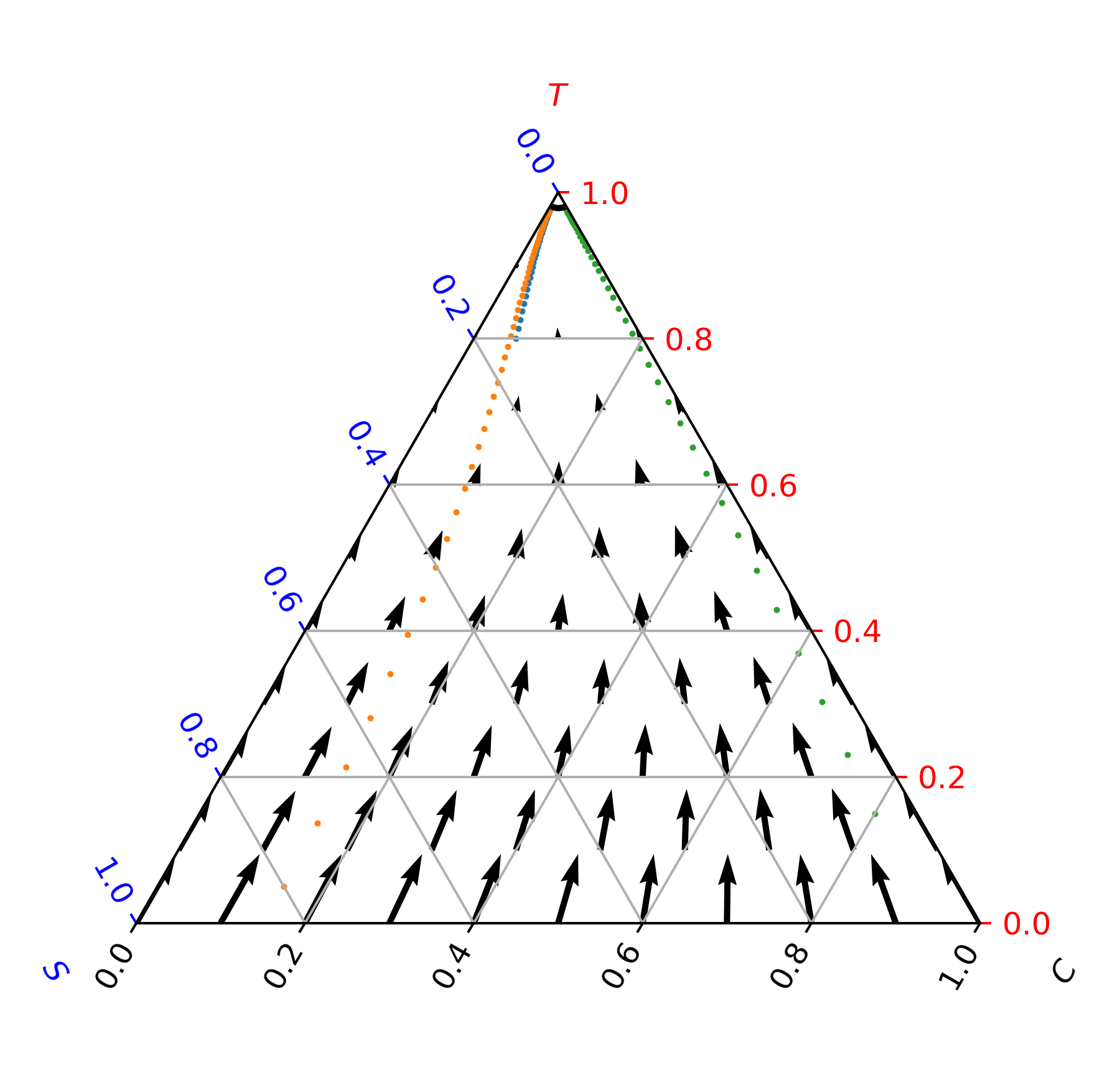}
        \caption{$\mu = 0.2$, $\gamma = 0.1$}\label{fig:x1}
    \end{subfigure}%
    \\
    \begin{subfigure}{0.333\textwidth}
    \centering
        \includegraphics[width =\textwidth]{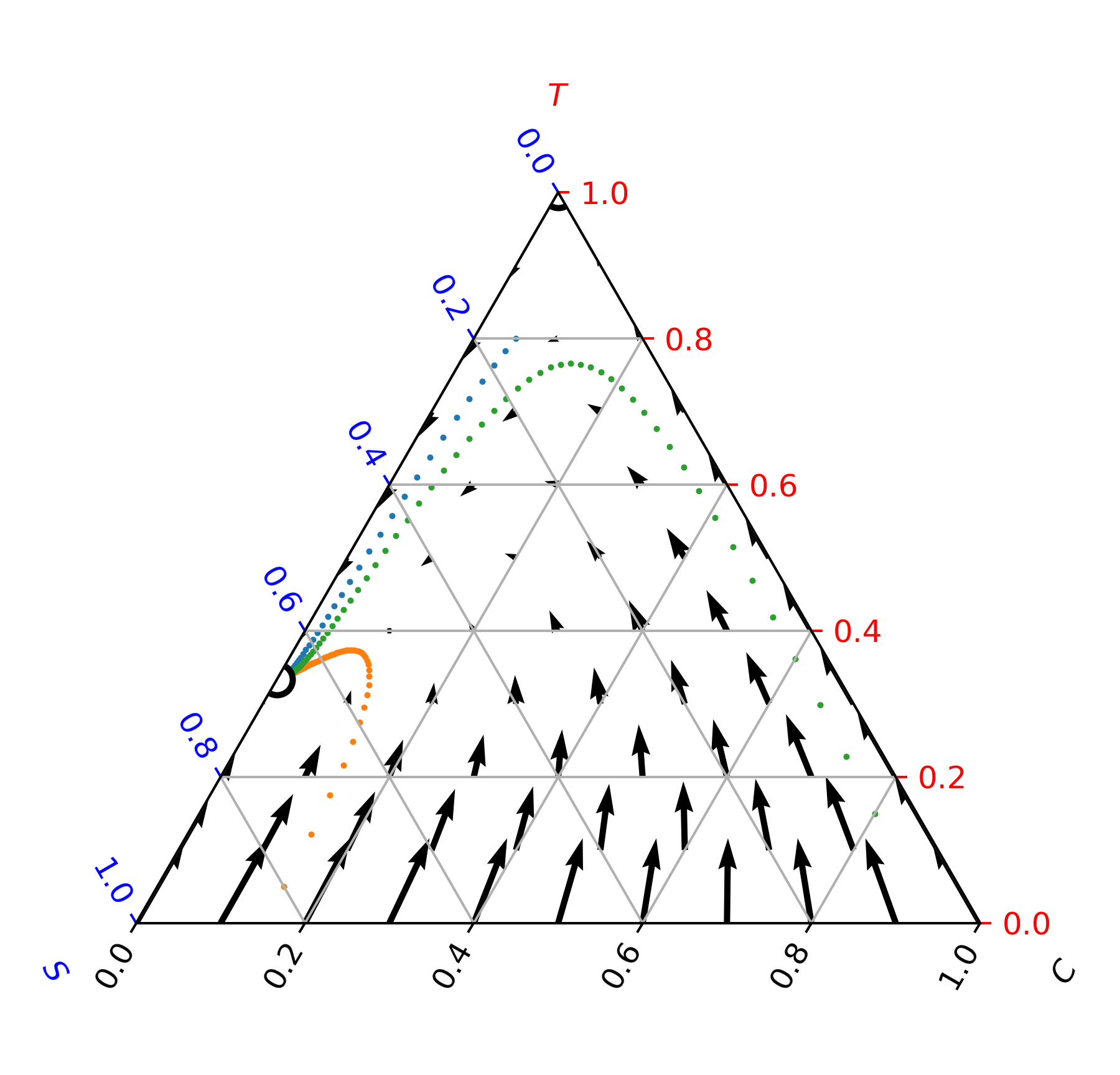}
        \caption{$\mu = 0.2$, $\gamma = 0.6$}\label{fig:x2}
    \end{subfigure}%
        \begin{subfigure}{0.333\textwidth}
    \centering
        \includegraphics[width = \textwidth]{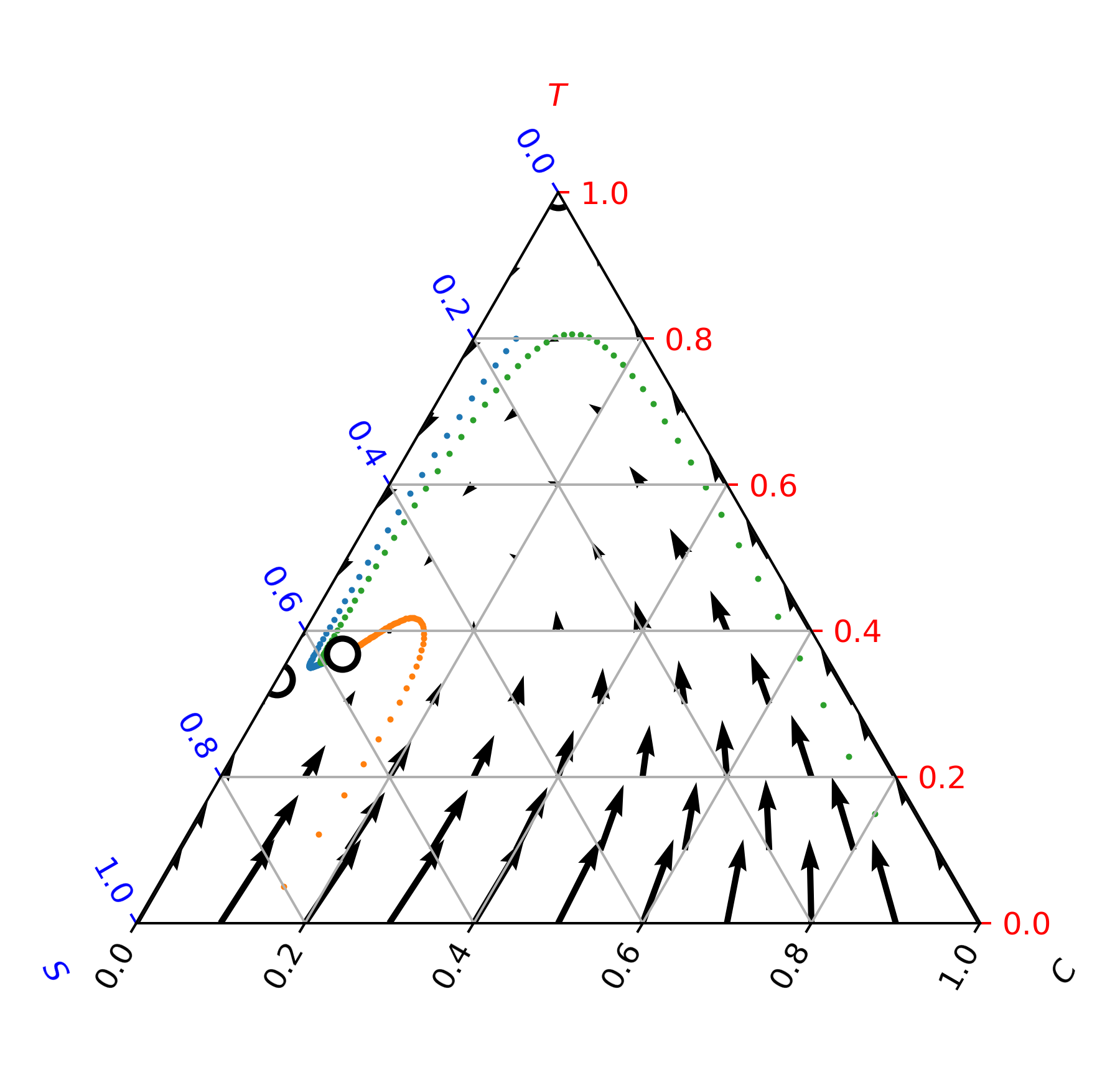}
        \caption{$\mu = 0.35$, $\gamma = 0.6$}\label{fig:x3node}
    \end{subfigure}%
    \begin{subfigure}{0.333\textwidth}
    \centering
        \includegraphics[width =\textwidth]{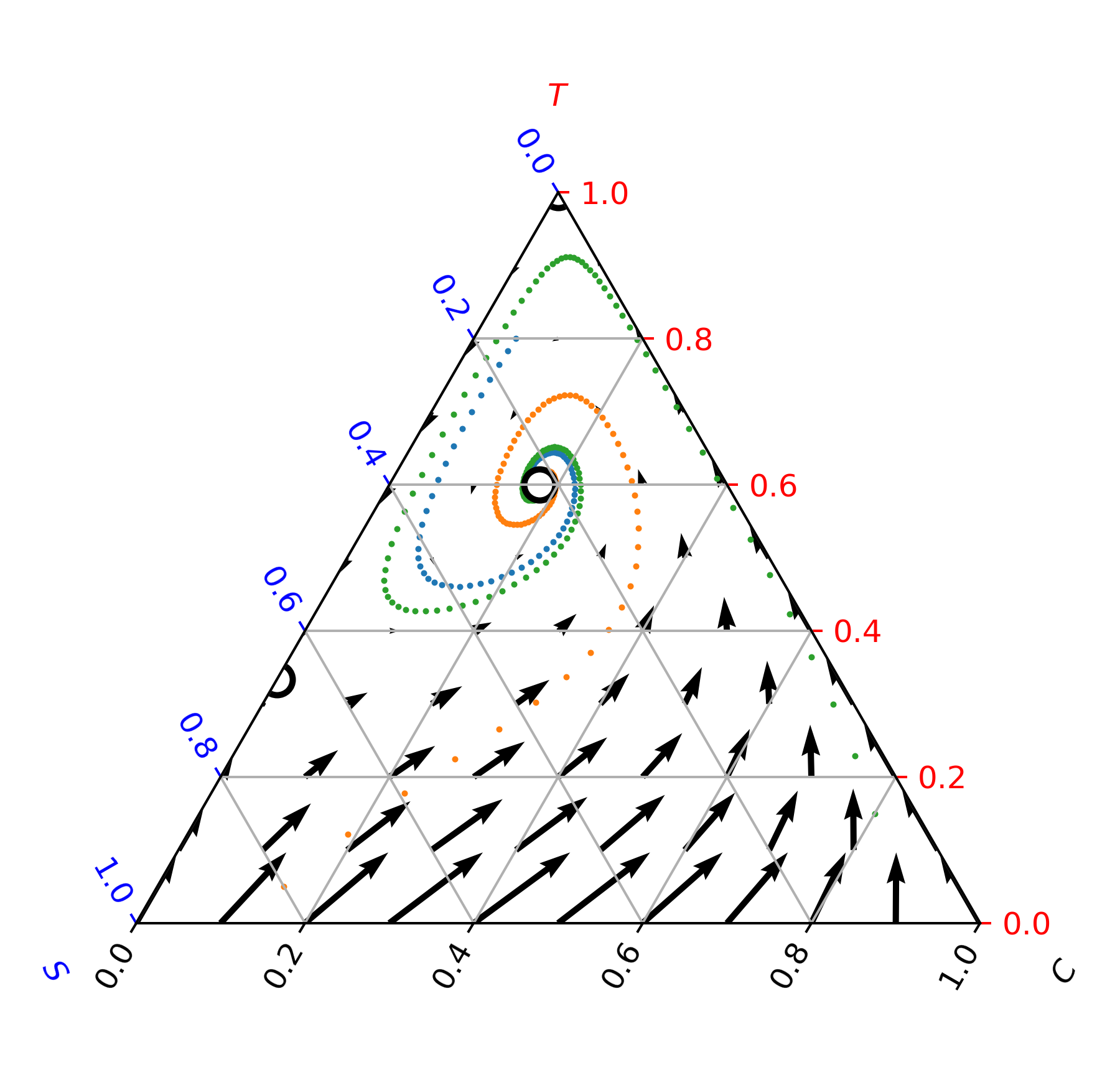}
        \caption{$\mu = 0.9$, $\gamma = 0.6$}\label{fig:x3spiral}
    \end{subfigure}%
     \caption{Phase plane for the stability and type of stationary points, assuming that $\nu<0.5$. In each region the stationary point in the label is the only \textit{stable} point (its type is also noted), while the remaining stationary points are saddle points. Ternary flow diagrams are shown illustrating the flow in each of the regions of the phase plane ($\Lambda = \nu = 0.2$).} 
    \end{figure}

For a fixed value of $\nu<0.5$ and various values of $\gamma, \mu$, the classification of the three stationary points is depicted in Figure~\ref{fig:Stability}. We observe that there is only one stable stationary point per parameter configuration. Flow diagrams (in ternary projection) are depicted in Figures~\ref{fig:x1}-\ref{fig:x3spiral}, illustrating the nature of the fixed points in each of the regions identified in the phase plane of Figure~\ref{fig:Stability}. Along with the flow, trajectories are scatter plotted, thus illustrating the expected dynamics from three points near the extremes.

%

\subsection{Numerical computation of fluid limit}
\begin{figure}[htb]
    \centering
    \begin{subfigure}{0.33\textwidth}
        \includegraphics[width = 0.99\textwidth]{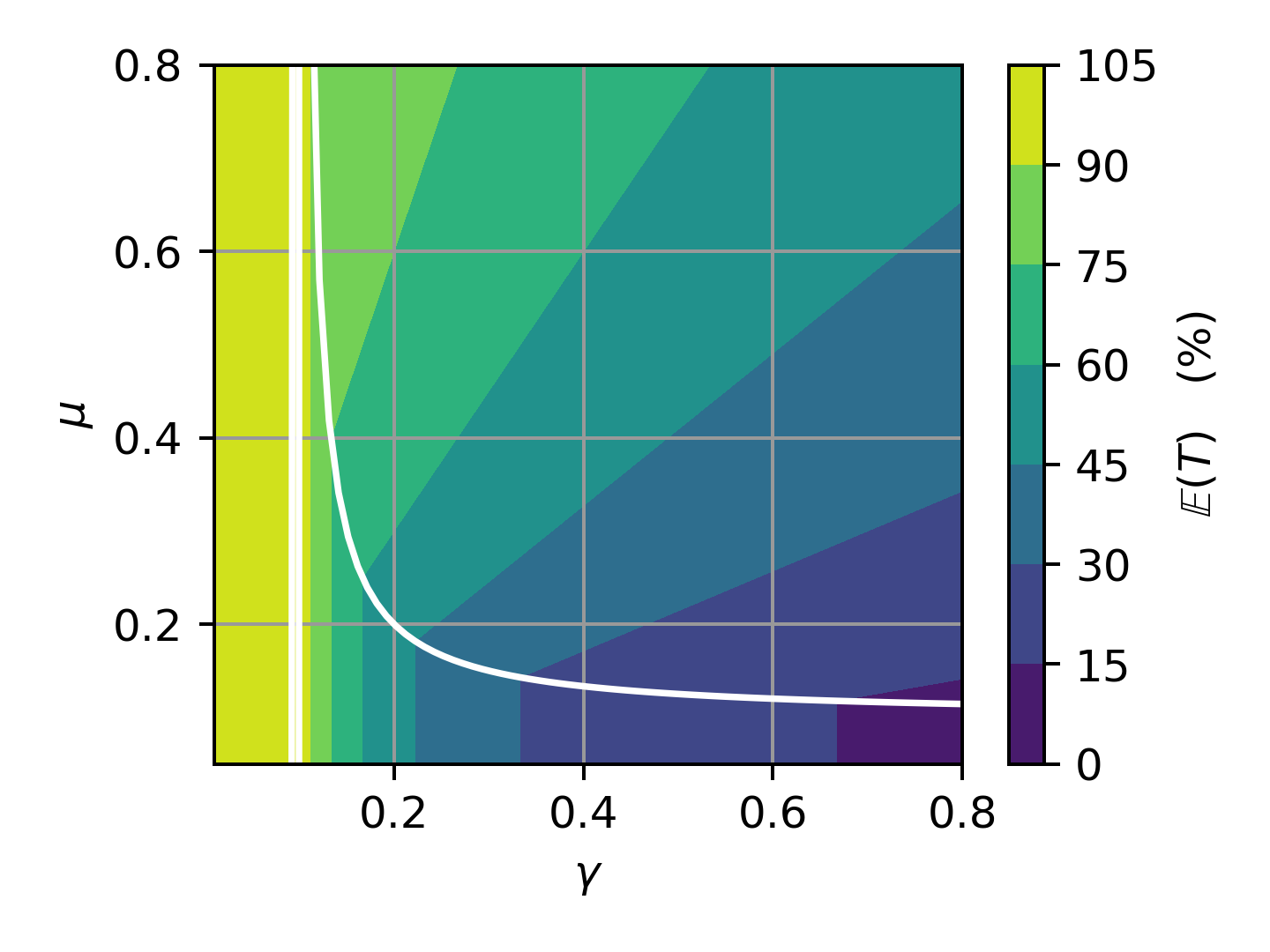}
        \caption{Trusters, $\nu=0.1$}
    \end{subfigure}%
    \begin{subfigure}{0.33\textwidth}
        \includegraphics[width = 0.99\textwidth]{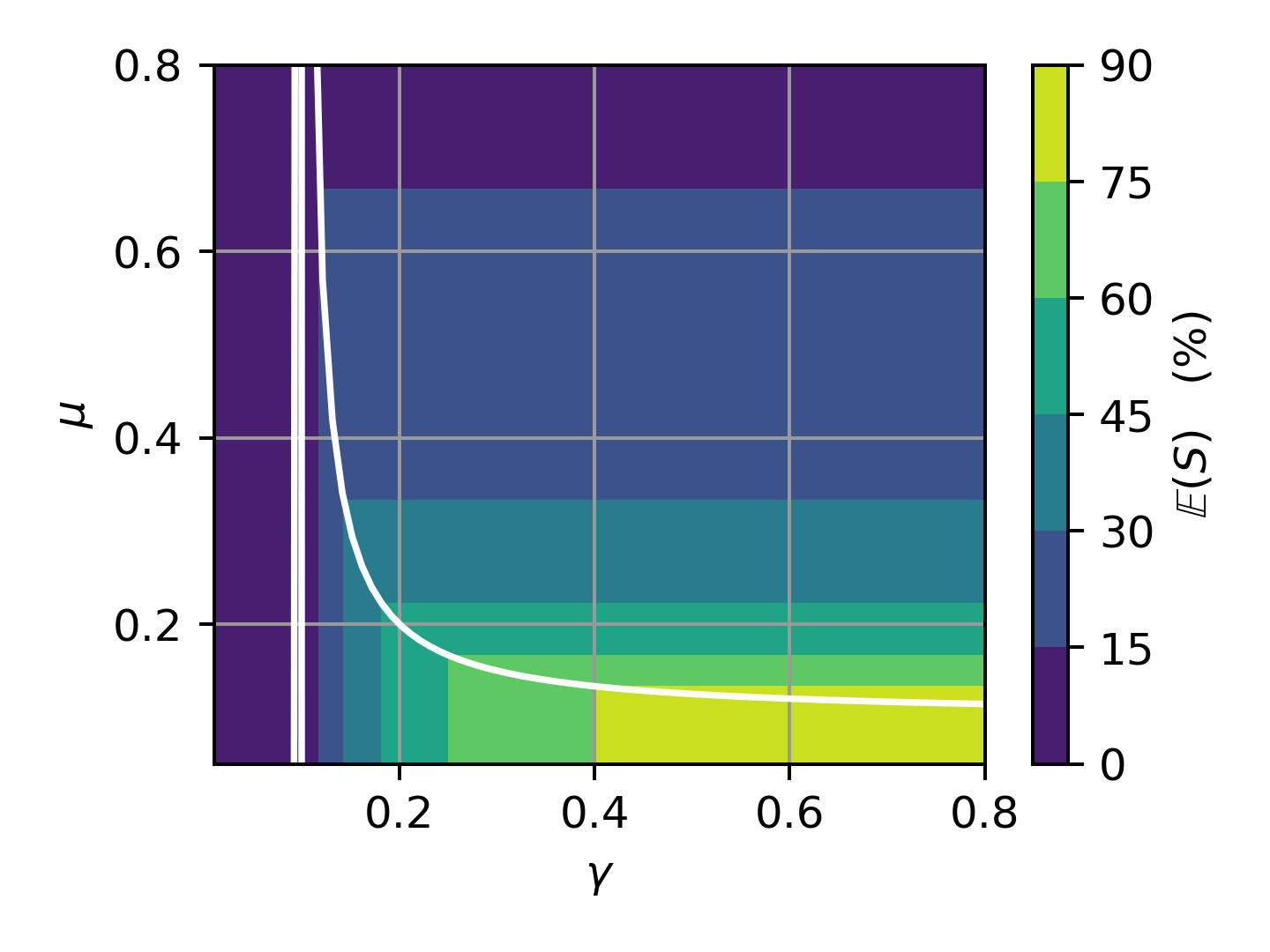}
        \caption{Skeptics, $\nu=0.1$}
    \end{subfigure}%
    \begin{subfigure}{0.33\textwidth}
        \includegraphics[width = 0.99\textwidth]{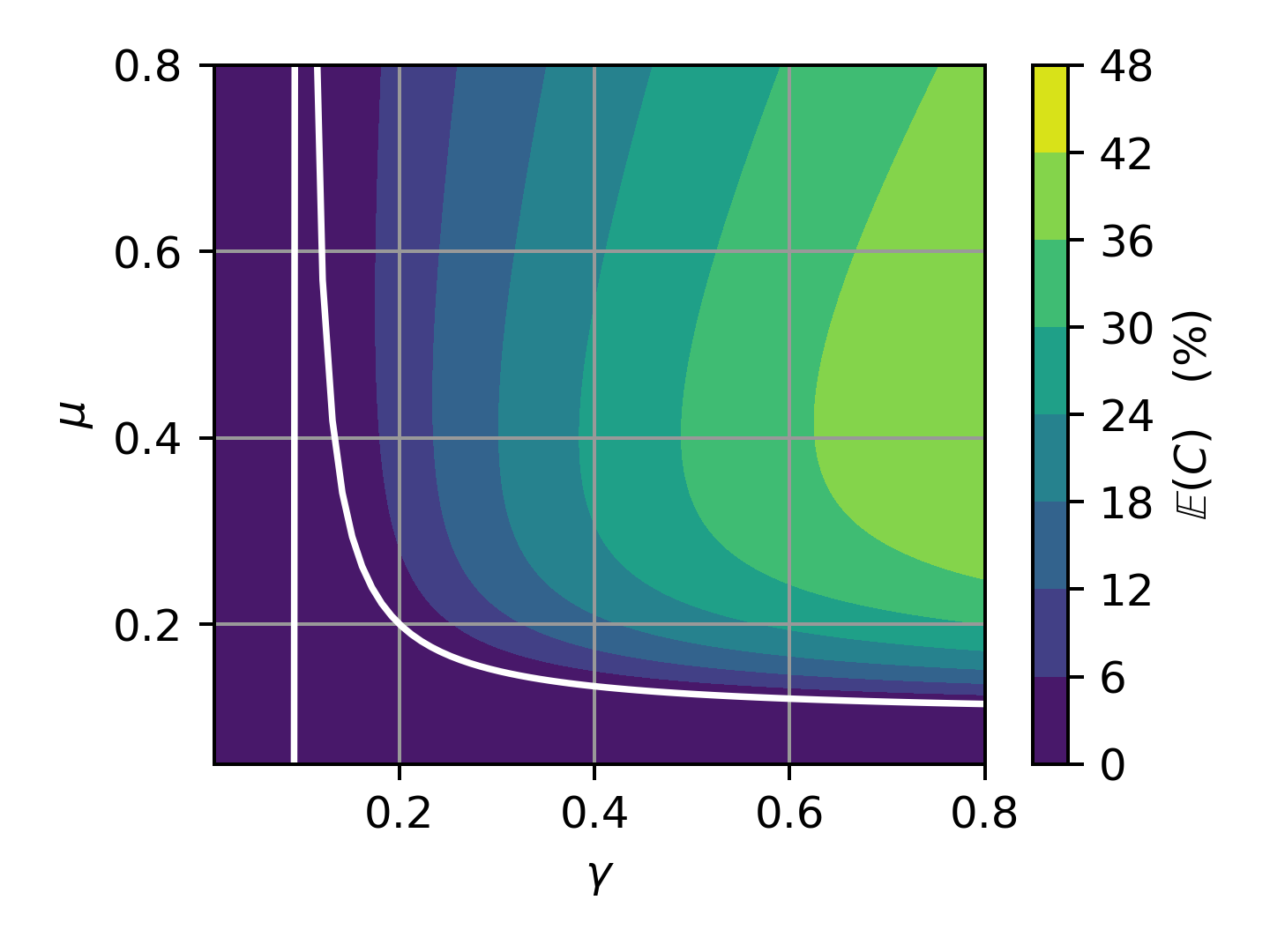}
        \caption{Doubters, $\nu=0.1$}\label{subfig:C_nu10}
    \end{subfigure}%
    \\
       \begin{subfigure}{0.33\textwidth}
        \includegraphics[width = 0.99\textwidth]{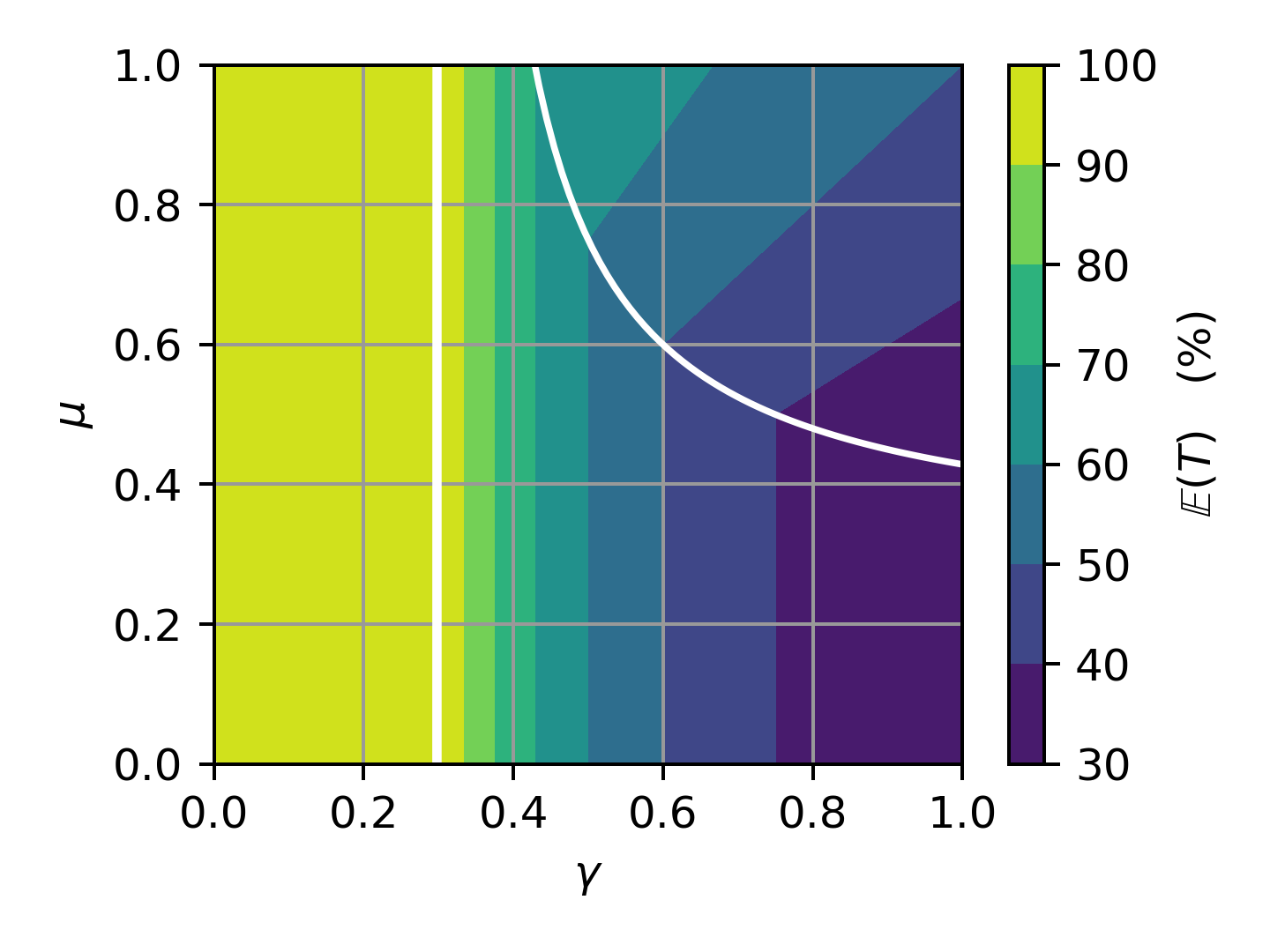}
        \caption{Trusters, $\nu=0.3$}
    \end{subfigure}%
    \begin{subfigure}{0.33\textwidth}
        \includegraphics[width = 0.99\textwidth]{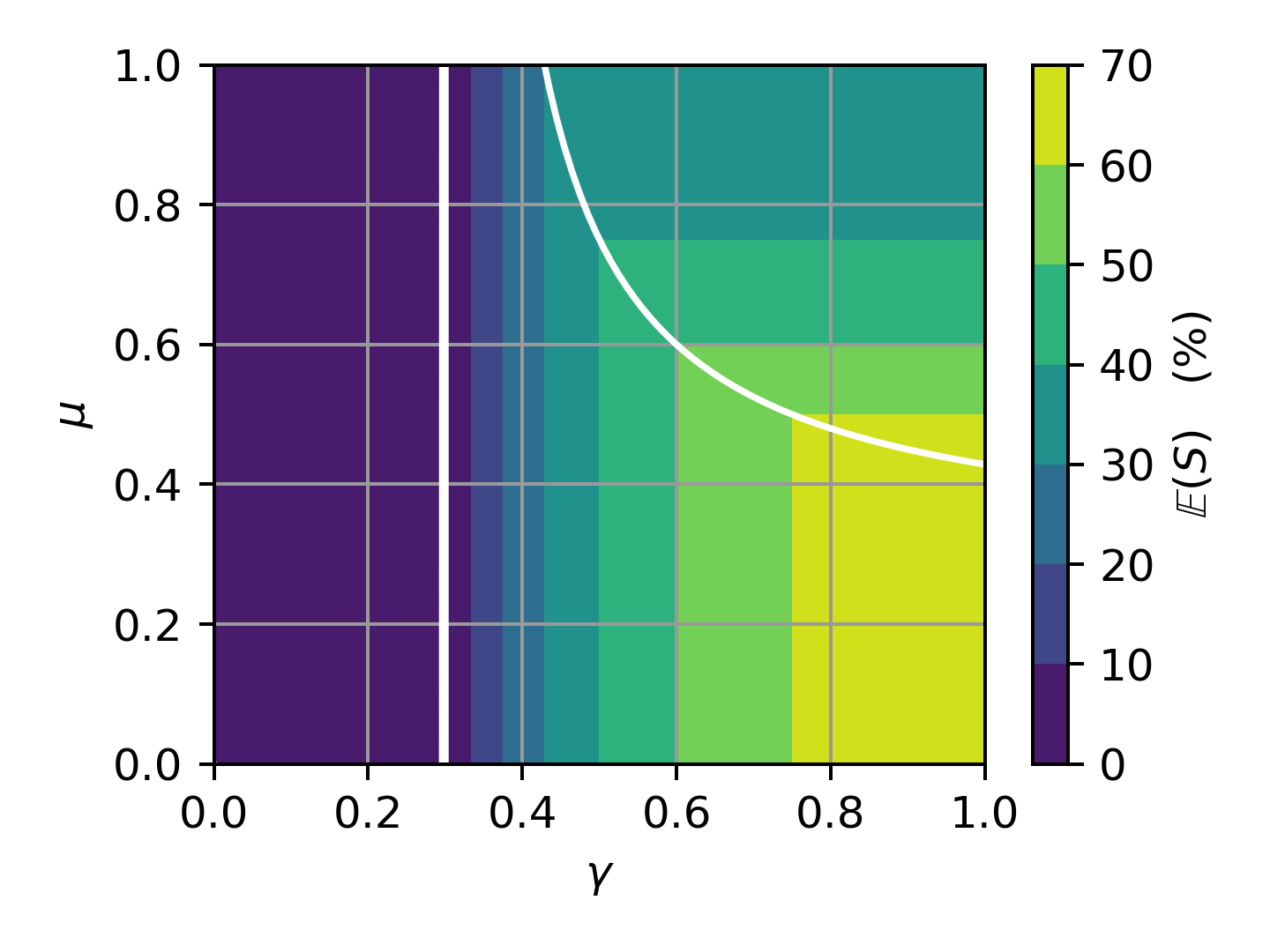}
        \caption{Skeptics, $\nu=0.3$}
    \end{subfigure}%
    \begin{subfigure}{0.33\textwidth}
        \includegraphics[width = 0.99\textwidth]{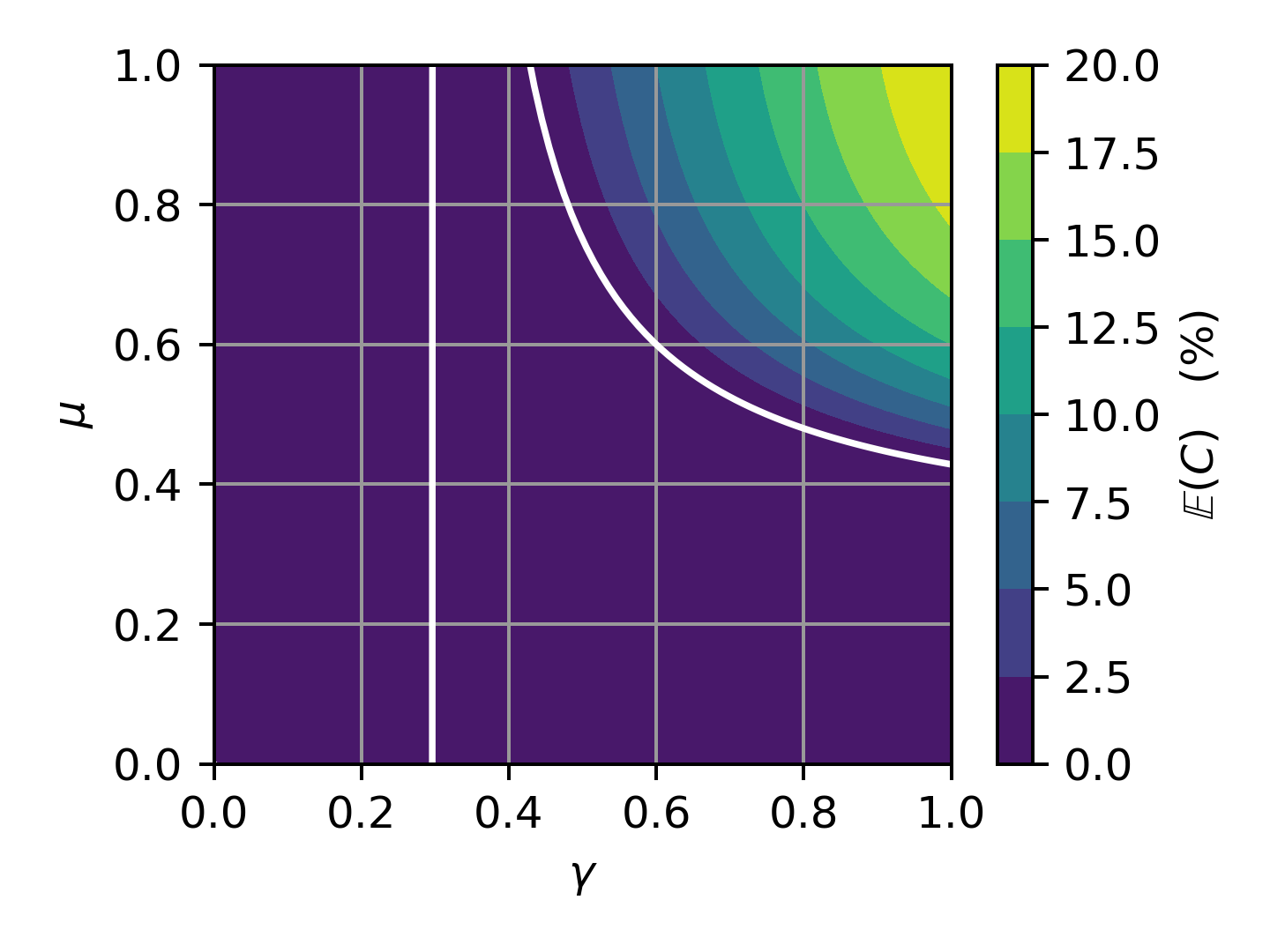}
        \caption{Doubters, $\nu=0.3$}
    \end{subfigure}%
    \caption{The fluid approximation results in heatmap format ($\nu=0.1, 0.3$). The white lines indicate the phase transition between the stability of stationary points. We see that as $\nu$ increases the range of parameter values for which $C$ does not die out shrinks.}
    \label{fig:heatmaps}
\end{figure}

We start by examining the fluid limit to identify which regions are of particular interest. The long-term expected number of $T_f,S_f$ and $C_f$ is plotted in Figure~\ref{fig:heatmaps} for a range of parameter values $\gamma,\mu$ and $\nu \in\{ 0.1,0.3\}$. The values plotted correspond to the stable (as discussed in \S\ref{sec:steady}) point from the set $\{\bm{X}_1, \bm{X}_2\bm{X}_3\}$ under the parameter value combination. 

We notice in the figure and by the dependence of stationary point $\bm{X}_3$ on the value of $\nu$ that a lower birth and death rate (greater life expectancy) facilitates the existence of the group who have lost their trust completely ($C_f(t)>0$ for all $t\geqslant 0$). 

We also see that as $\gamma$ increases (the net rate of transition from trusting to skepticism), the expected number of trusters decreases, while the number of skeptics increases. If $\mu$ is great enough, then as $\gamma$ increases, the expected proportion of the $C_f$ group also increases. 

Furthermore, in Figure~\ref{subfig:C_nu10} we see that the relationship between $C_f$ and $\mu$ is not monotone. In particular, there is a point after which increasing $\mu$ (the rate at which group $C_f$ persuades individuals from group $S_f$), the expected number of $C_f$ in steady state \textit{decreases}. 

Important to note is that the size of the $S_f$ group may be very small in steady state $\bm{X}_3$. The expression $S_f=\nu/\mu$ may become small, because in order for $\bm{X}_3$ to be stable we require the denominator $\mu$ to be sufficiently large, or, more precisely,
\[\mu>\frac{\gamma^{3/2}\sqrt{\nu} +\gamma\nu}{2(\gamma-\nu)},\] while at the same time the numerator $\nu$ should be sufficiently small in order to ensure $\gamma>\nu$. For the system in the fluid limit this is not an issue, while for the stochastic system in the pre-limit, a small $S$ group might die out by random fluctuation and therewith also ensure the extinction of the $C$ group. To investigate the size of this effect we consider the diffusion approximation as well as the Gillespie simulation of the pre-limit system in \S\ref{sec:diffusion_simulation}.

\subsection{Pre-limit stochastic analysis}\label{sec:diffusion_simulation}
There are multiple ways to analyze the fluctuations of the stochastic model from the fluid path. In this subsection, we focus on the diffusion approximation as well as the Gillespie type simulation. 

As pointed out, we are in the setting of~\cite[Theorem~8.2]{Kurtz1981}, thus we can approximate the stochastic system by its fluid limit mean plus a noise term which behaves as a multivariate Gaussian process. We recall that  the covariances between the three  components, at time $t\geqslant 0$, are given by the solution to
\begin{equation}\label{eq:var}
    \diff{{\bm V}(t)}{t} = \nabla \bm{F}(\bm{x}(t)) {\bm V}(t) + {\bm V}(t) \left(\nabla \bm{F}(\bm{x}(t))\right)^\top + H\Sigma(\bm{x}(t))\times \left(H\Sigma(\bm{x}(t))\right)^\top;
\end{equation}
for our compartmental model, we have that the matrices $H$ and $\Sigma(t)$ are specified by
\begin{equation}
    H=\begin{bmatrix}
        1 & -1 & 1 & 0 & -1 & 0 & 0\\ 
        0 & 1 & -1 & -1 & 0 & -1 & 0\\
        0 & 0 & 0 & 1 & 0 & 0 & -1
    \end{bmatrix},
\end{equation}
and 
\begin{equation}\label{eq:Sigmat}
    \Sigma(t) = \text{diag}\left( \sqrt{\Lambda}, \sqrt{\alpha T_f(t)S_f(t)}, \sqrt{\beta T_f(t)S_f(t)}, \sqrt{\mu S_f(t)C_f(t)}, \sqrt{\nu T_f(t)}, \sqrt{\nu S_f(t)}, \sqrt{\nu C_f(t)} \right).
\end{equation}
Notice that for the diffusion approximation, unlike the fluid approximation, we cannot track only $\gamma=\alpha-\beta$, because $\alpha$ and $\beta$ appear individually in (\ref{eq:Sigmat}).

For completeness we also specify $\nabla \bm{F}(\bm{x}(t))$:
\begin{equation}
    \nabla \bm{F}(\bm{x}(t)) = \begin{bmatrix}
        S_f(t) (\beta-\alpha)-\nu & T_f(t)(\beta-\alpha)& 0\\
        S_f(t)(\alpha-\beta) &T_f(t)(\alpha-\beta)-C_f(t)\mu -\nu &\mu S_f(t)\\
        0 &C_f(t)\mu &S_f(t) \mu-\nu
    \end{bmatrix}.
\end{equation}
Given the result of the fluid path, being readily attained by a numerical initial value problem solver (such as those available in the Python packages SciPy and NumPy), Equation (\ref{eq:var}) may be solved using by a standard routine such as \texttt{odeintw}~\cite{Weckesser2015} (a wrapper around the numerical solvers in SciPy).

It is possible to simulate the system using the ordinary differential equation (\ref{eq:fluid_de}) in combination with Gaussian noise generated by the stochastic differential equation (\ref{eq:dif_sde}), though when the system is stable this can be avoided by instead solving directly for expected values, variances and covariances. These may be combined with a Python3 (or similar) implementation of a multi-variate Gaussian distribution to determine quantities of interest, such as the probability of a compartment being empty (or close to it). In this section we present the results of Gillespie simulation as well as approximations based on the fluid/diffusion limit.

\subsubsection{Expected outcome}
\begin{figure}[htb]
    \centering
   \begin{subfigure}{0.32\textwidth}
       \centering
        \includegraphics[width = 0.99\textwidth]{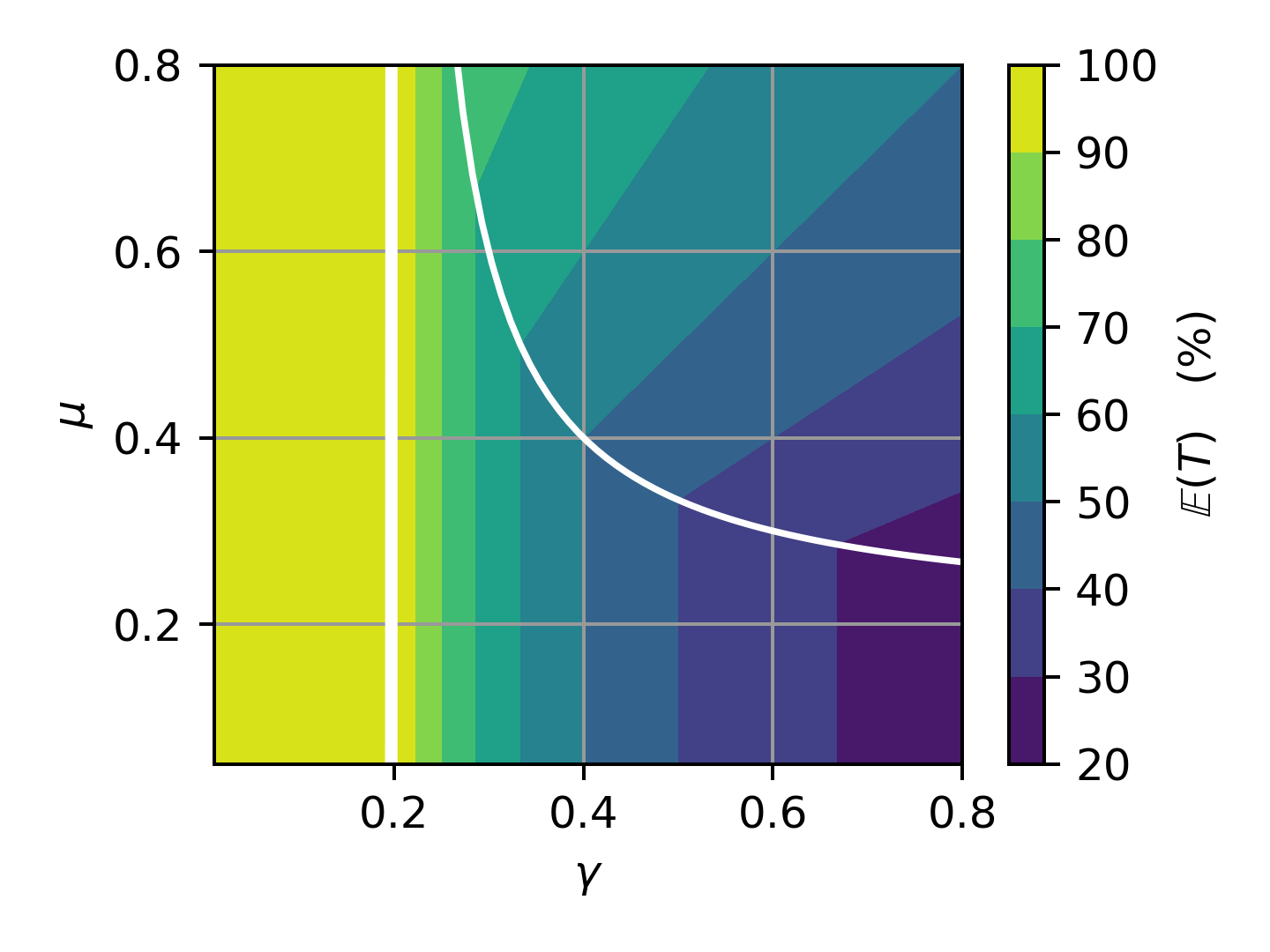}
        \caption{$T$ Fluid limit}
    \end{subfigure}%
    \begin{subfigure}{0.32\textwidth}
        \centering
        \includegraphics[width = 0.99\textwidth]{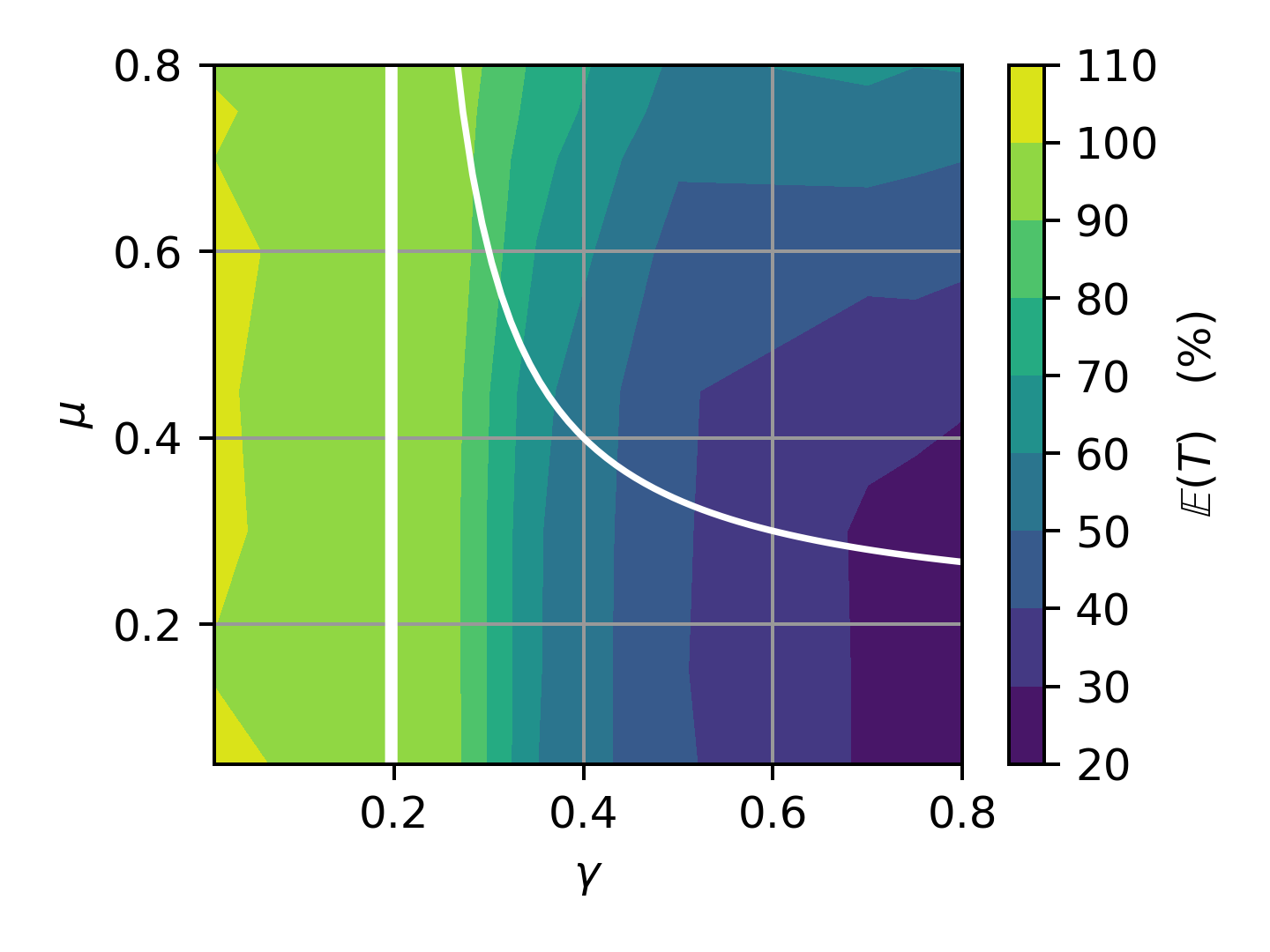}
        \caption{$T$ Simulation $N=100$}
    \end{subfigure}%
    \begin{subfigure}{0.32\textwidth}
        \centering
        \includegraphics[width = 0.99\textwidth]{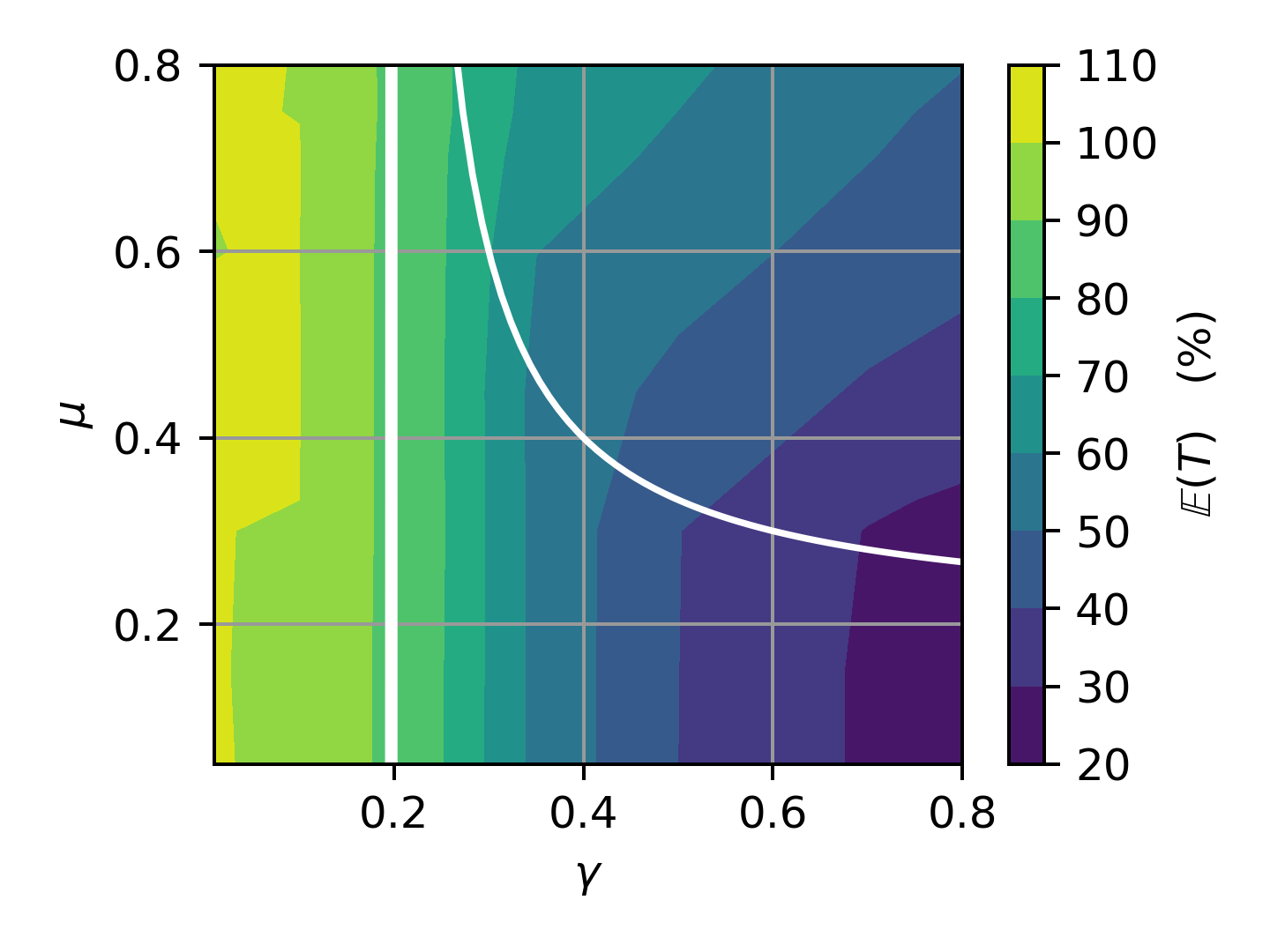}
        \caption{$T$ Simulation $N=1000$}
    \end{subfigure}%
        \caption{Proportion of individuals in the $T$ compartment in steady state for the fluid limit as well as expected value of $T/N$ at the termination of the simulation given by the Gillespie simulation with $\nu=0.2$. The white lines indicate the phase transition between the stability of stationary points.}
    \label{fig:heatmaps_simT}
    \end{figure}

\begin{figure}[htb]
    \centering
       \begin{subfigure}{0.32\textwidth}
           \centering
        \includegraphics[width = 0.99\textwidth]{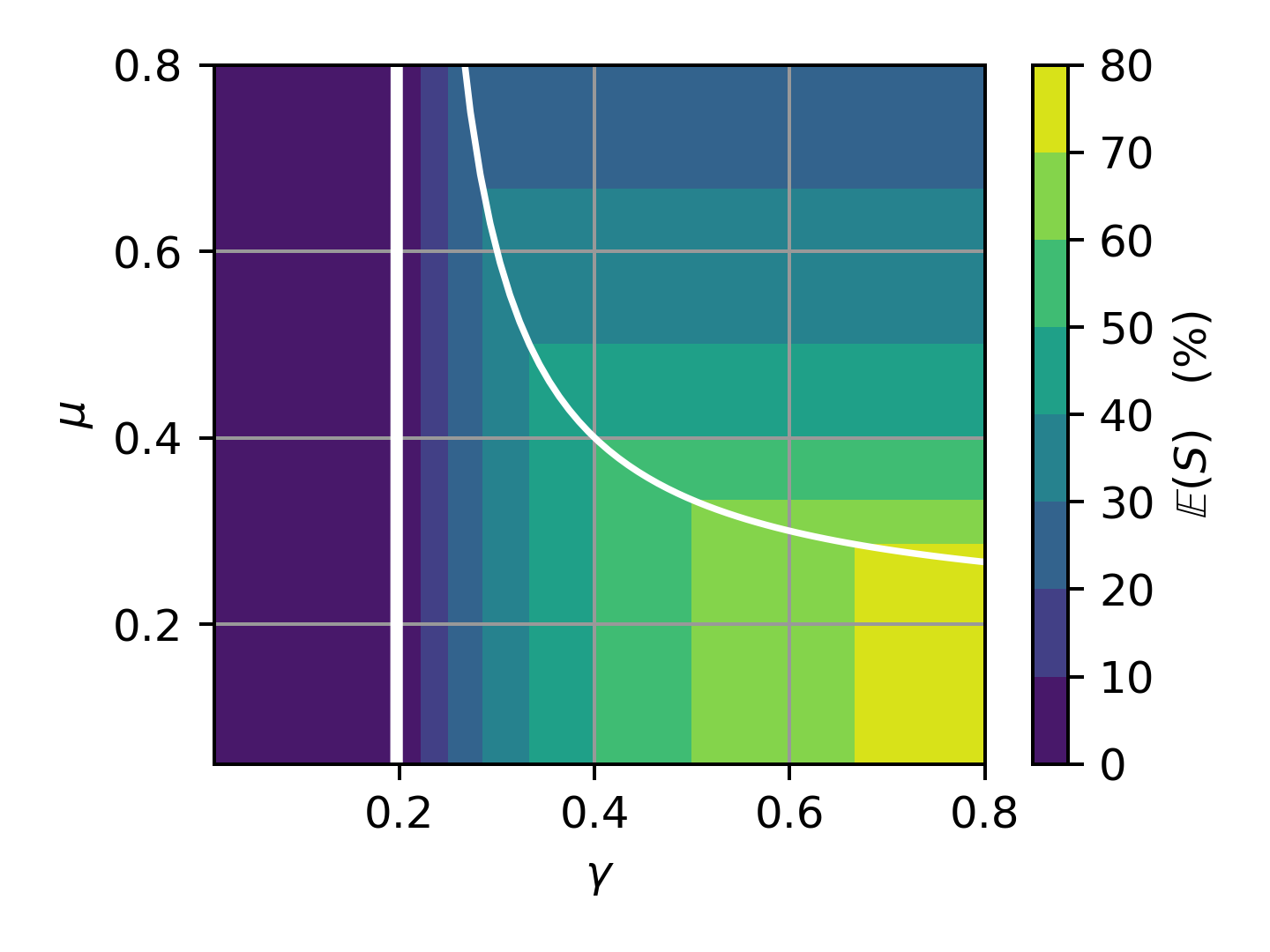}
        \caption{$S$ Fluid limit}
    \end{subfigure}%
    \begin{subfigure}{0.32\textwidth}
        \centering
        \includegraphics[width = 0.99\textwidth]{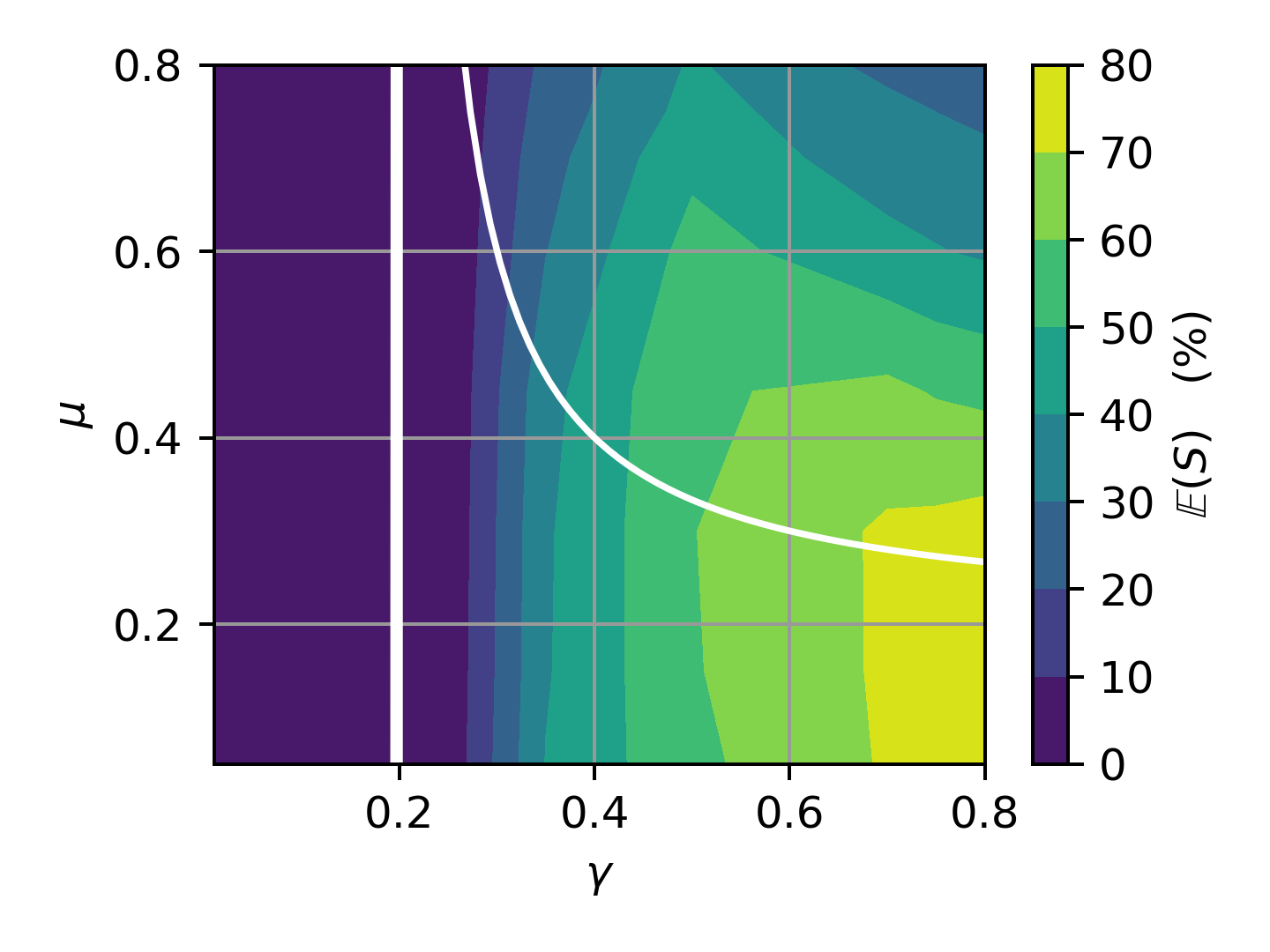}
        \caption{$S$ Simulation $N=100$}
    \end{subfigure}%
    \begin{subfigure}{0.32\textwidth}
        \centering
        \includegraphics[width = 0.99\textwidth]{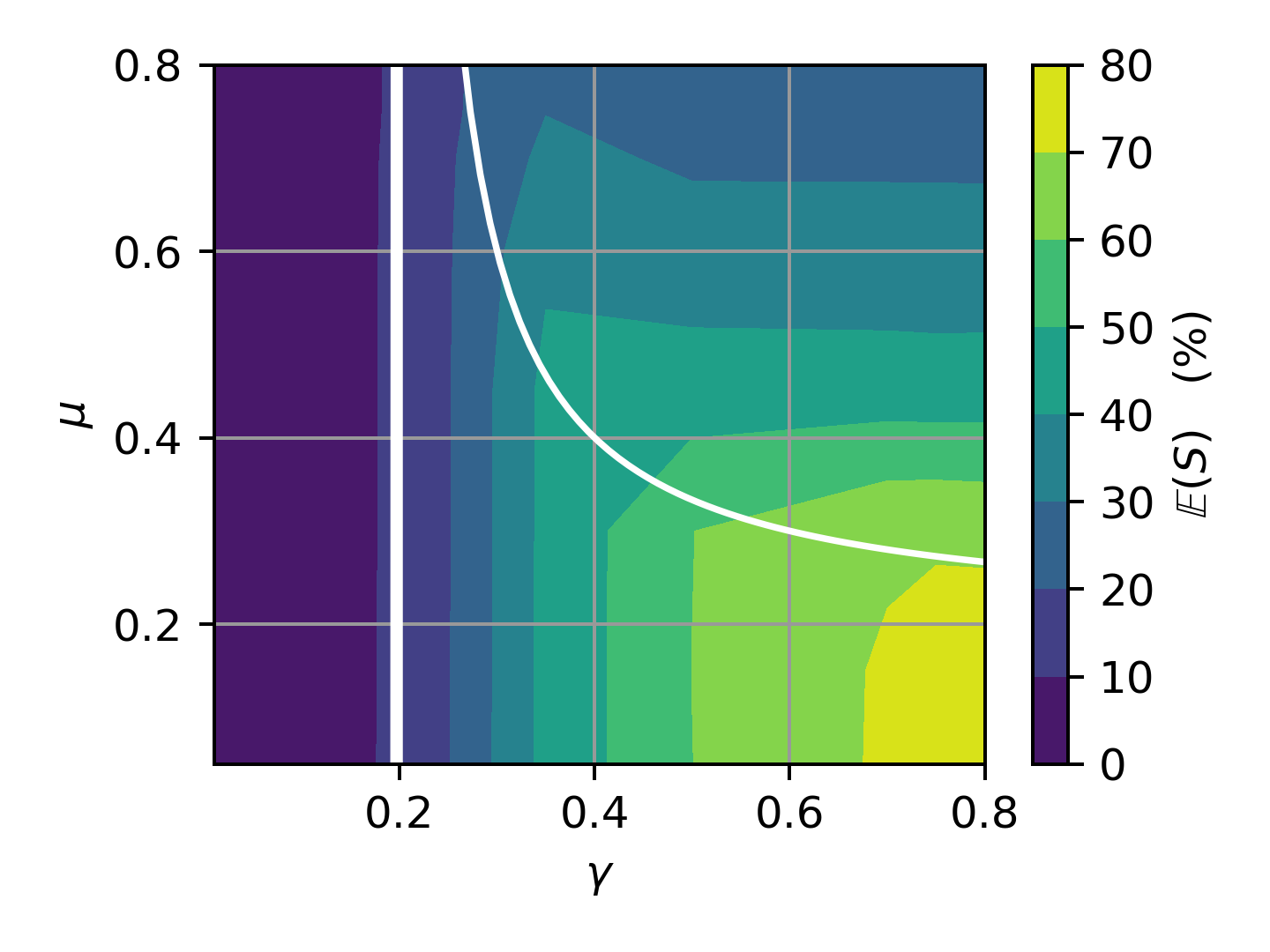}
        \caption{$S$ Simulation $N=1000$}
    \end{subfigure}%
        \caption{Proportion of individuals in the $S$ compartment in steady state for the fluid limit as well as expected value of $S/N$ at the termination of the simulation given by the Gillespie simulation with $\nu=0.2$. The white lines indicate the phase transition between the stability of stationary points.}
    \label{fig:heatmaps_simS}
    \end{figure}

\begin{figure}[htb]
    \centering
       \begin{subfigure}{0.32\textwidth}
           \centering
        \includegraphics[width = 0.99\textwidth]{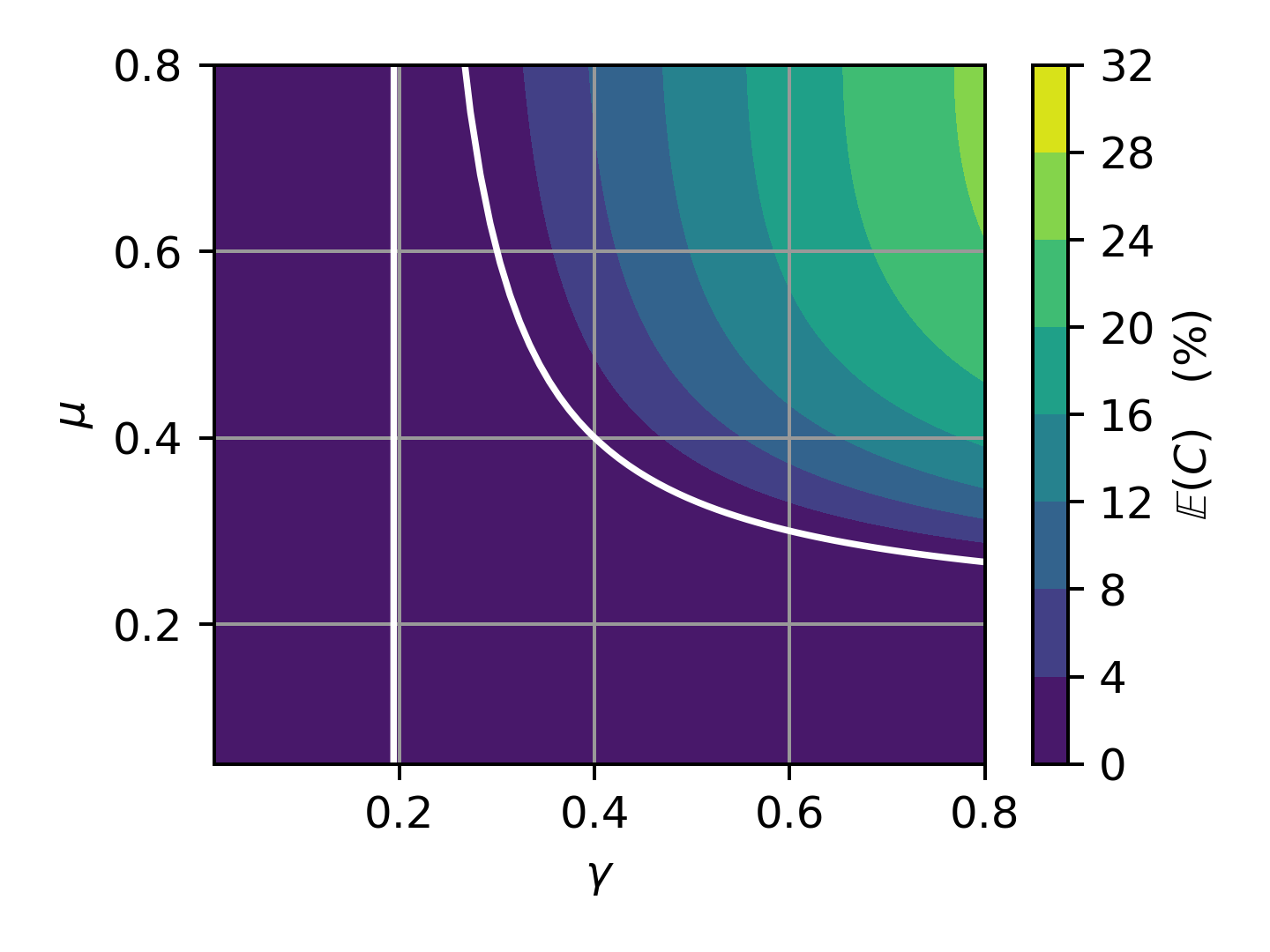}
        \caption{$C$ Fluid limit}\label{subfig:fl_c}
    \end{subfigure}%
    \begin{subfigure}{0.32\textwidth}
        \centering
        \includegraphics[width = 0.99\textwidth]{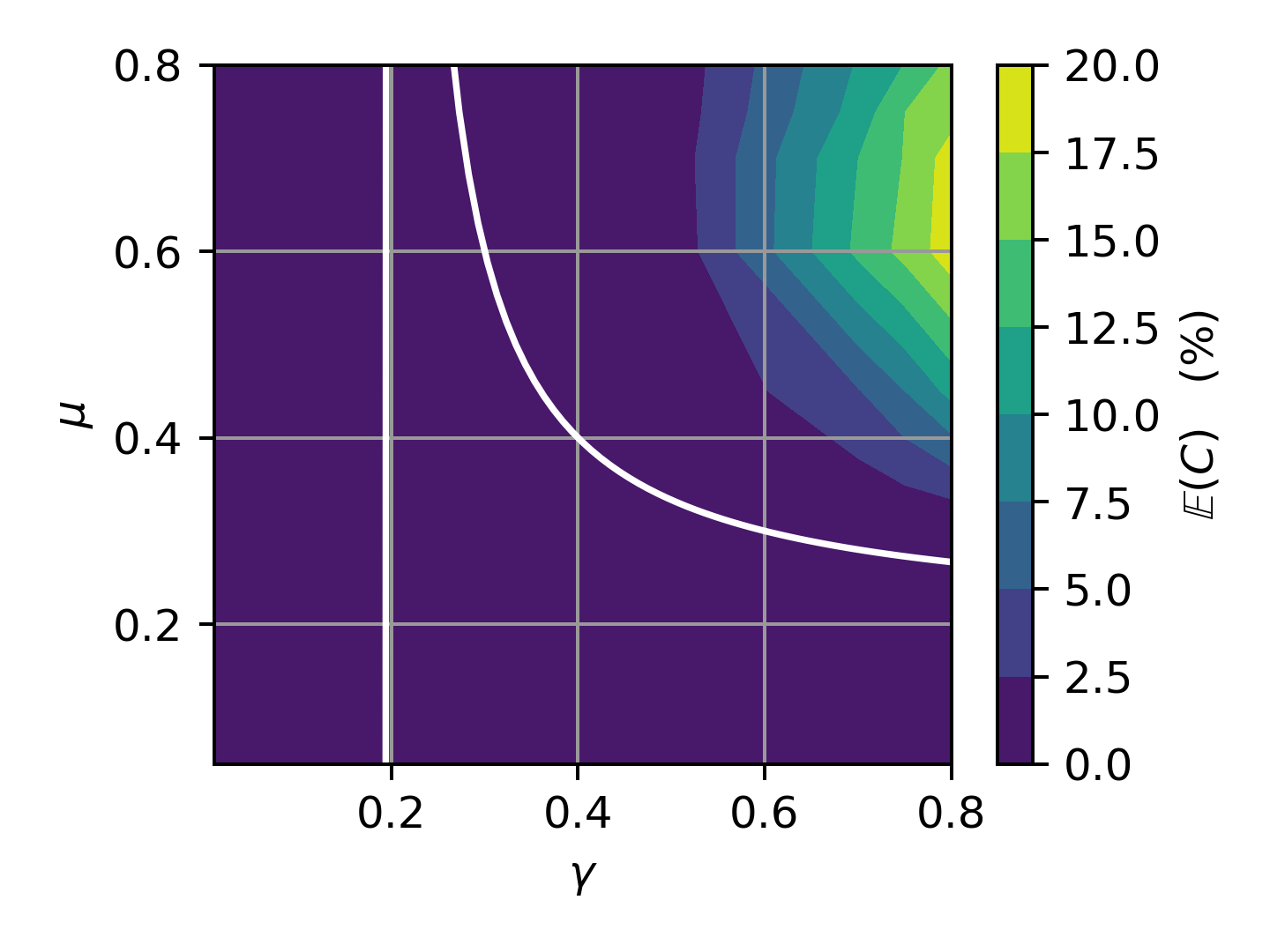}
        \caption{$C$ Simulation $N=100$}\label{subfig:sim_c100}
    \end{subfigure}%
    \begin{subfigure}{0.32\textwidth}
        \centering
        \includegraphics[width = 0.99\textwidth]{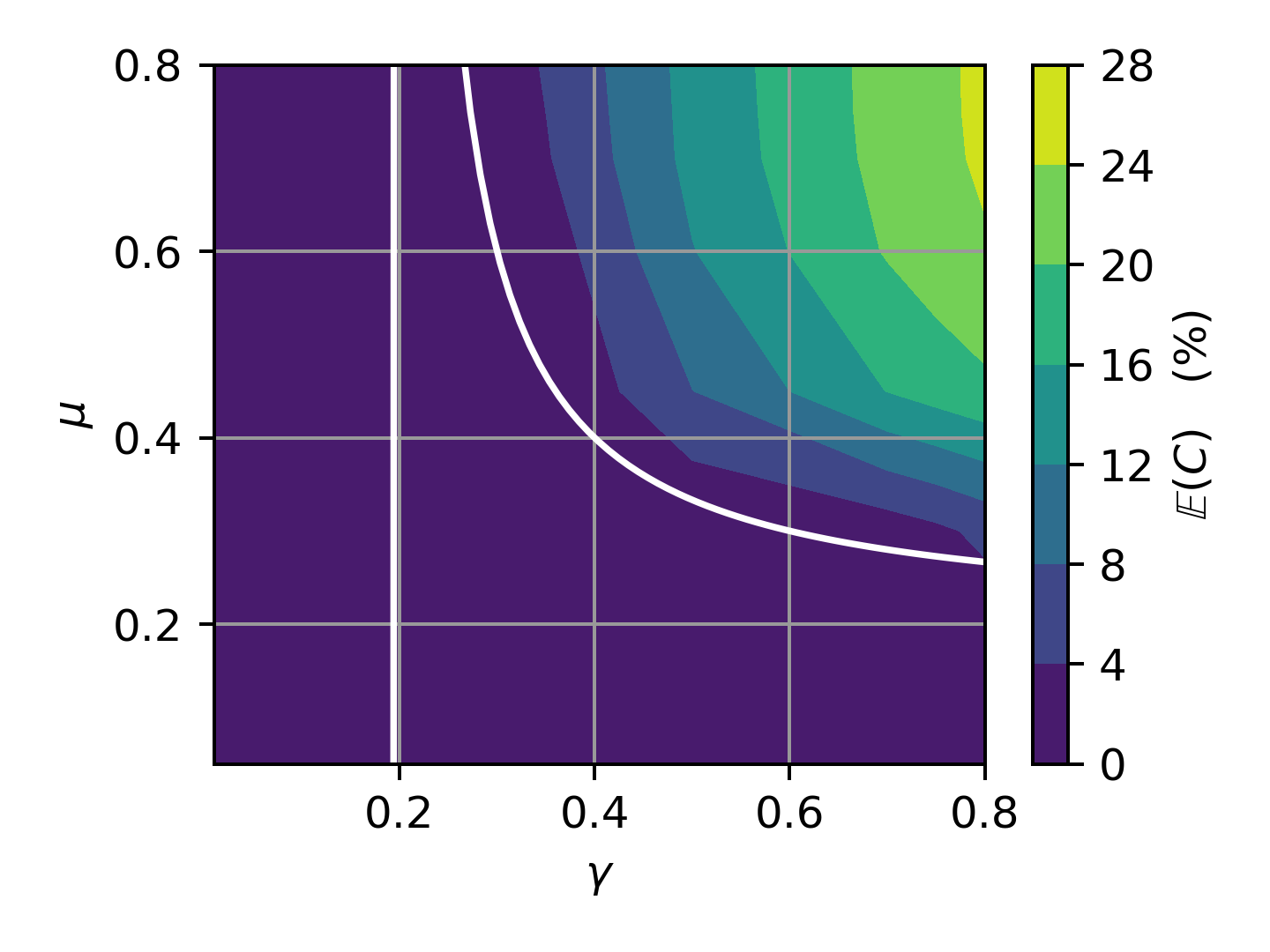}
        \caption{$C$ Simulation $N=1000$}\label{subfig:sim_c1000}
    \end{subfigure}%
    \caption{Proportion of individuals in the $C$ compartment in steady state for the fluid limit as well as expected value of $C/N$ at the termination of the simulation given by the Gillespie simulation with $\nu=0.2$. The white lines indicate the phase transition between the stability of stationary points.}
    \label{fig:heatmaps_simC}
\end{figure}

In Figures~\ref{fig:heatmaps_simT}--\ref{fig:heatmaps_simC} we show the long-term expected proportion of individuals in each group for the fluid limit as well as the results obtained by Gillespie simulation using $N=100,$ and $N=1000$ at $\nu=0.2$. In each plot, the $\gamma$ and $\mu$ parameter values can be read off the $x$ and $y$ axis respectively, while the shading indicates the what percentage of the final population is in the compartment in question. For the simulated results we give the proportion of final individuals per group compared to the initial total number of individuals. This means that it is possible for the $T$ group to have more than $100\%$ in $\bm{X}_1$ due to the intrinsic fluctuations of the stochastic system.

By comparing Figures~\ref{subfig:fl_c},~\ref{subfig:sim_c100}, and~\ref{subfig:sim_c1000} we can see that the simulations persistently predict a lower proportion of individuals who have lost their trust (in the region where this is not expected to die-out). This effect becomes less from $N=100$ to $N=1000$, and by the convergence theorem of Kurtz~\cite{Kurtz1981}, as $N\to\infty$ this discrepancy will disappear. Recall that if the $S$ or the $C$ compartment is empty at some time $t_0$ then this stays true for all $t>t_0$. In the simulation this happens at positive probability (even when $C_f(t)>0$ and $S_f(t)>0$ for all $t$) which means there are simulation runs with dynamics far from their expectation in the fluid limit. 

We also notice that in general the sharp phase transition in the fluid limit are smoothed out in the stochastic system. In particular the transition between $\bm{X}_1$ and $\bm{X}_2$ is much less sharp. This is because if the expected size of the $S$ group is small, it is likely that the fluctuations in the pre-limit model are enough to empty the $S$ compartment.

\subsubsection{Probability of skeptics and doubters dying out}\label{sec:probSkep_die}
We remind the reader that if the $S$ compartment dies out, so does the $C$ compartment. Thus we estimate the probability of $C$ dying out in the simulation by counting simulation runs in which \textit{only} $C$ dies out, and adding this to the number of runs in which $S$ dies out. By construction these events are mutually exclusive and so the sum of their probabilities is the probability the event that $C$ dies out. 

\begin{figure}[htb]
    \centering
    \begin{subfigure}{0.45\textwidth}
    \centering
    \vstretch{.9}{\includegraphics[width = 0.95\textwidth]{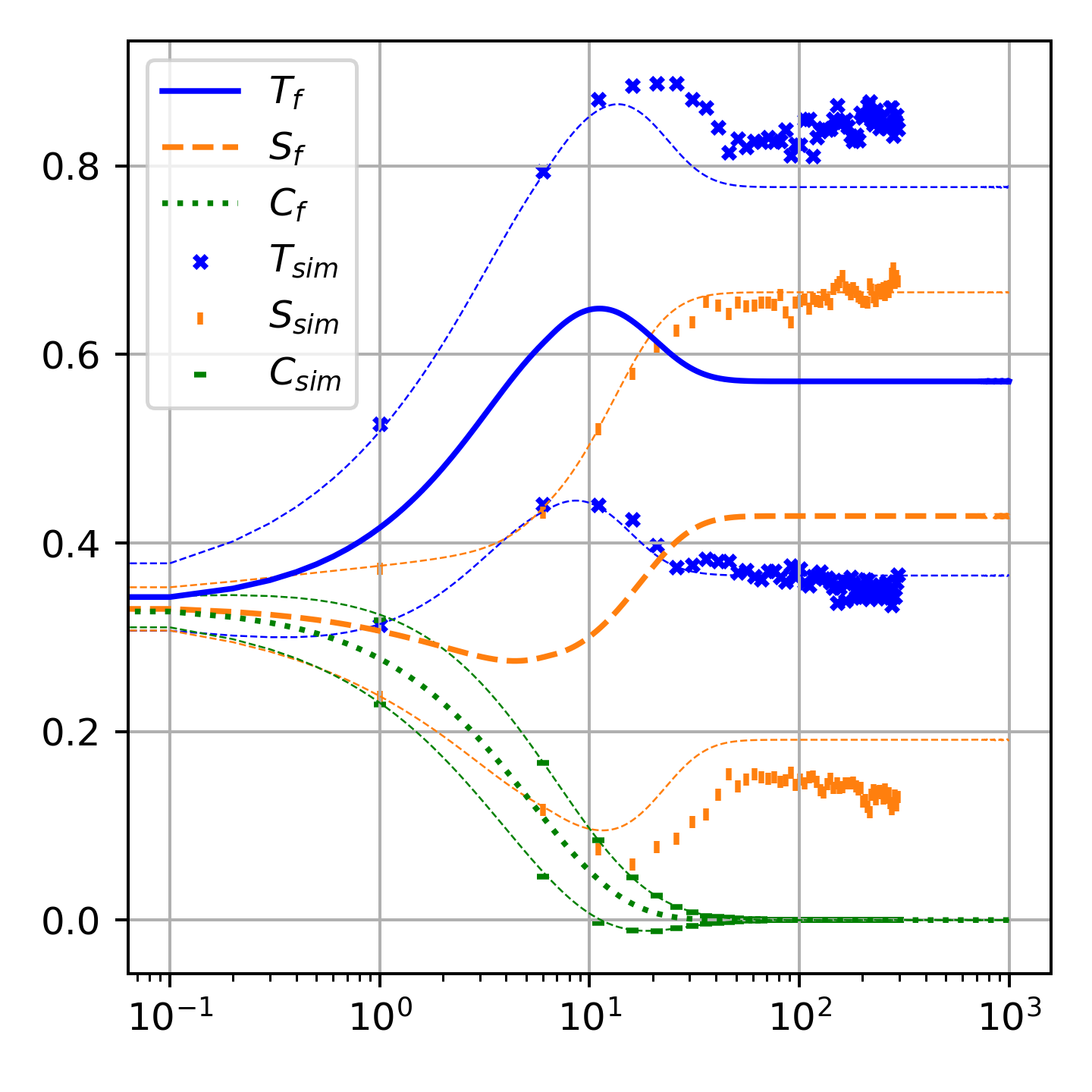}}
        \caption{$N=100$}\label{fig:sim_100}
    \end{subfigure}%
    \begin{subfigure}{0.45\textwidth}
    \centering
      \vstretch{.9}{\includegraphics[width =0.95\textwidth]{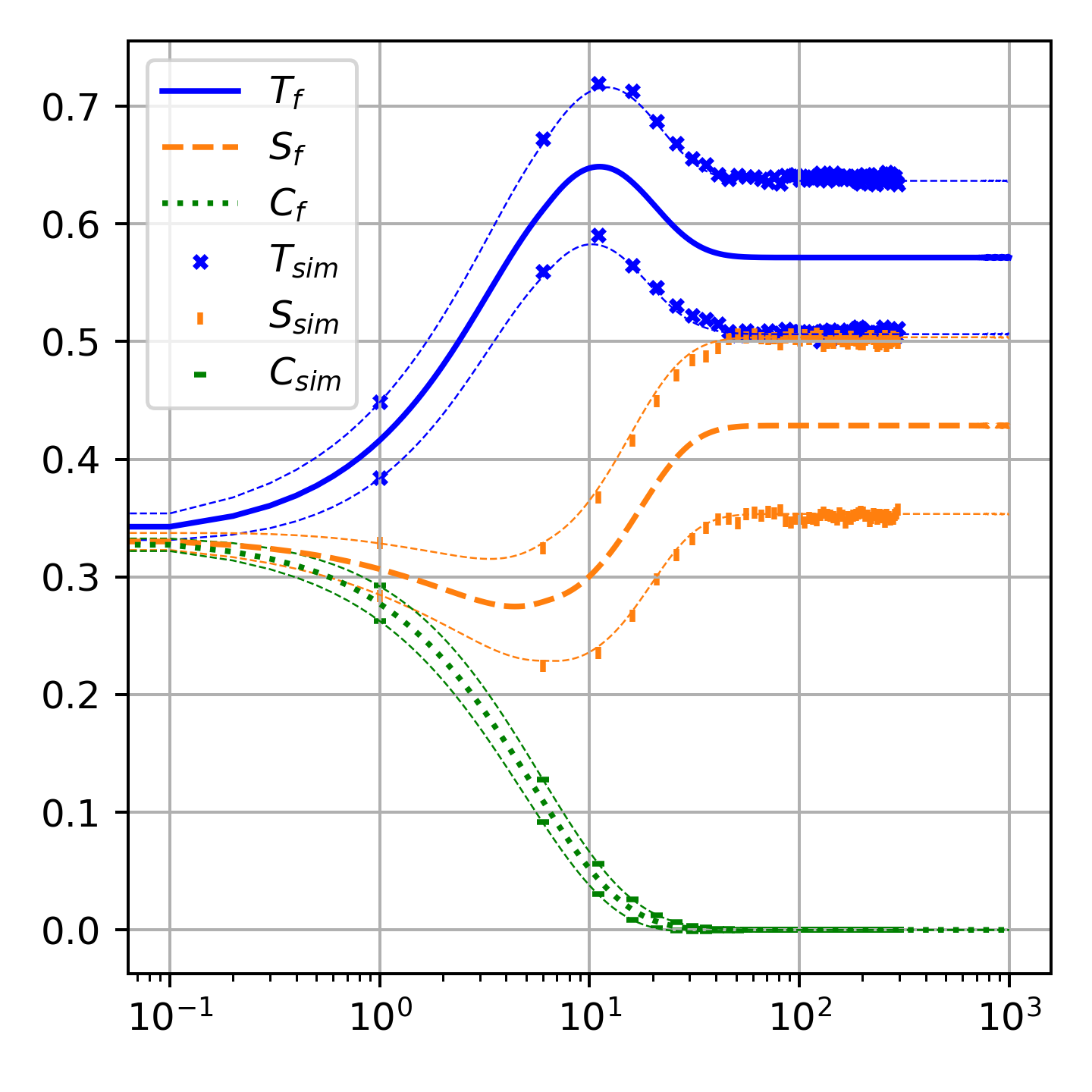}}
        \caption{$N=1000$}\label{fig:sim_1000}
    \end{subfigure}%
    \caption{Fluid and diffusion approximation with $\nu = 0.2, \mu = 0.05$, $\alpha=0.45$, and $\beta=0.1$ ($\gamma=0.35$) over time. Confidence intervals for the simulated system are also plotted.}
    \label{fig:well-behaved-approx}
\end{figure}

In Figure~\ref{fig:well-behaved-approx} we plot an illustration of the dynamics over time for the fluid and diffusion approximation as well as confidence intervals for the Gillespie simulation. For the simulation we use 500 iterations and the parameter settings are $\nu = 0.2$, $\alpha = 0.45$, $\beta=0.1$ ($\gamma= 0.35$), $\mu=0.05$, and $N=100,1000$. We see that the approximation gets better for greater $N$ (which is to be expected), but that even for $N=100$ the approximation is not far off. We note that in this parameter region, $C$ is expected to die out which means that this is also guaranteed in the stochastic model. 

\begin{figure}[htb]
    \centering
    \begin{subfigure}{0.45\textwidth}
    \centering
       \vstretch{.9}{\includegraphics[width = 0.95\textwidth]{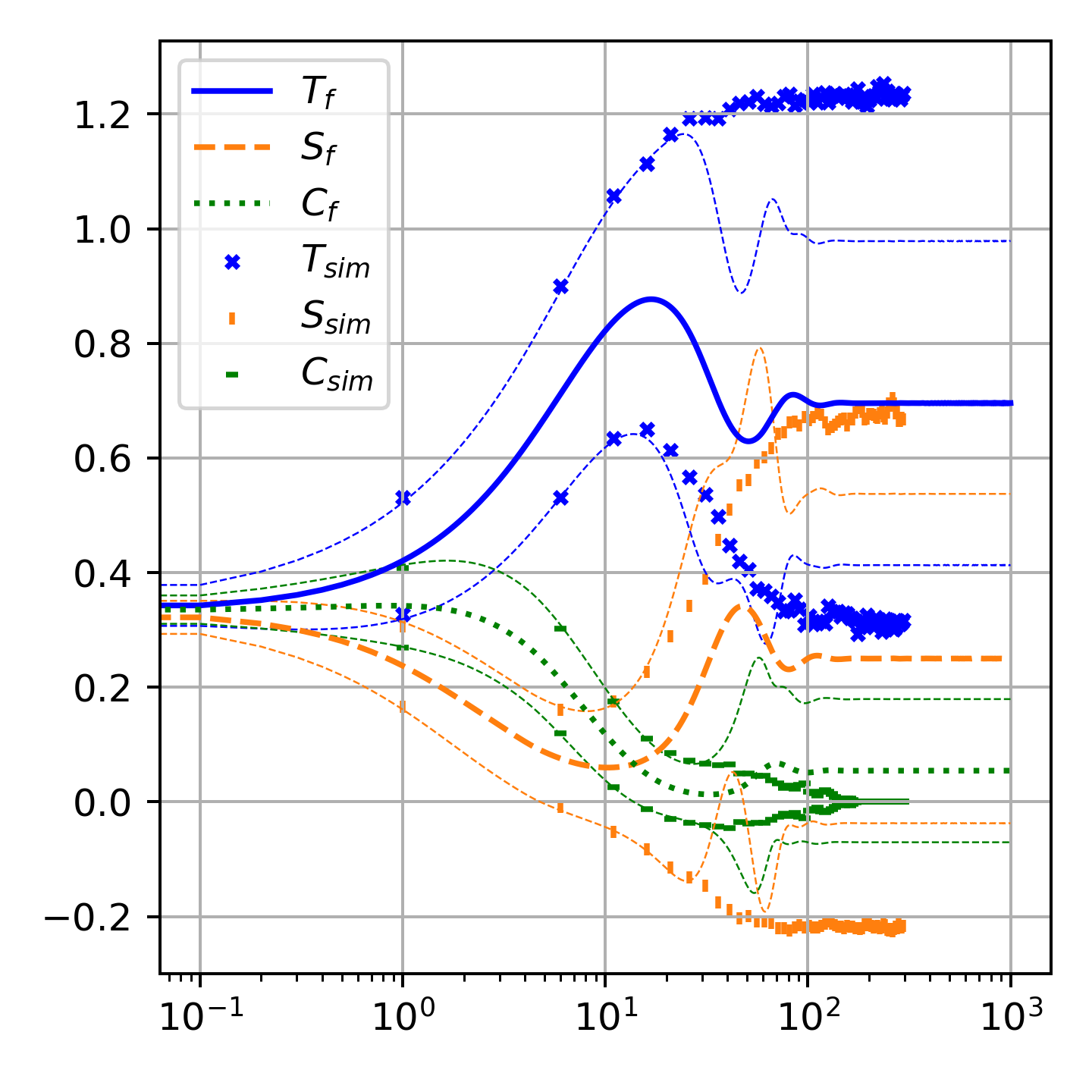}}
        \caption{$N=100$}\label{fig:sim_100_bad}
    \end{subfigure}%
    \begin{subfigure}{0.45\textwidth}
    \centering
        \vstretch{.9}{\includegraphics[width =0.95\textwidth]{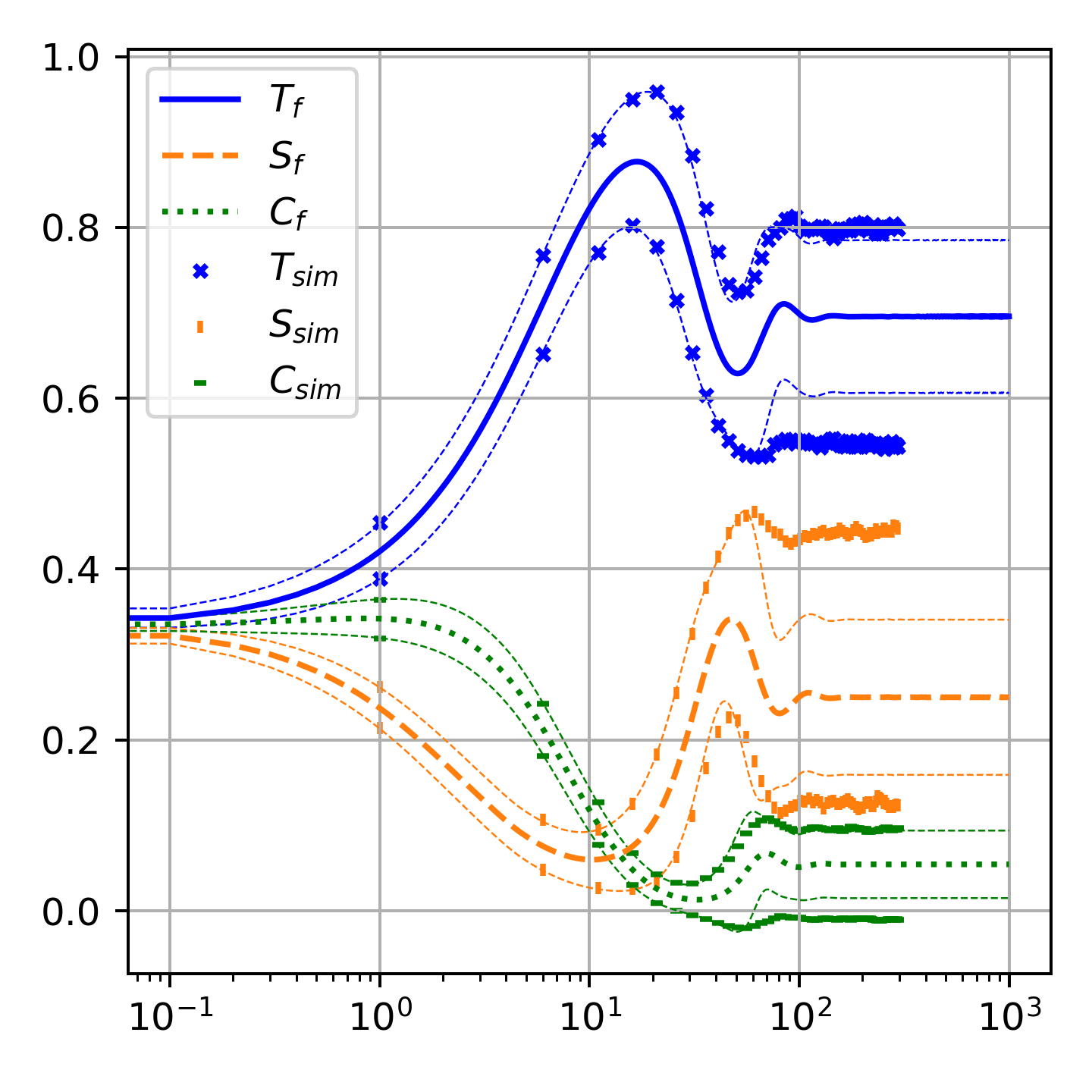}}
        \caption{$N=1000$}\label{fig:sim_1000_bad}
    \end{subfigure}%
    \caption{Fluid and diffusion approximation with $\nu = 0.2, \mu = 0.80$, $\alpha=0.45$, and $\beta=0.1$ ($\gamma=0.35$) over time. Confidence intervals for the simulated system are also plotted.}
    \label{fig:bad-behaved-approx}
\end{figure}

In Figure~\ref{fig:bad-behaved-approx} we plot the dynamics for the parameter settings $\nu = 0.2$, $\alpha = 0.45$, $\beta=0.1$ ($\gamma= 0.35$), $\mu=0.80$, and $N=100,1000$. In this case we see that in both cases ($N=100,1000$) the approximation underestimates the likelihood of $C$ dying out which results in the difference between the confidence intervals for the approximation and the simulation. In fact for $N=100$ all 500 simulation resulted in $C$ dying out.

The discrepancy between the simulation and the approximation indicates that in order to investigate the probability of the group without trust ($C$) dying out we are better served by the simulation than the approximation.

\begin{figure}[htb]
    \centering
    \begin{subfigure}{0.45\textwidth}
        \centering
\includegraphics[width = \textwidth]{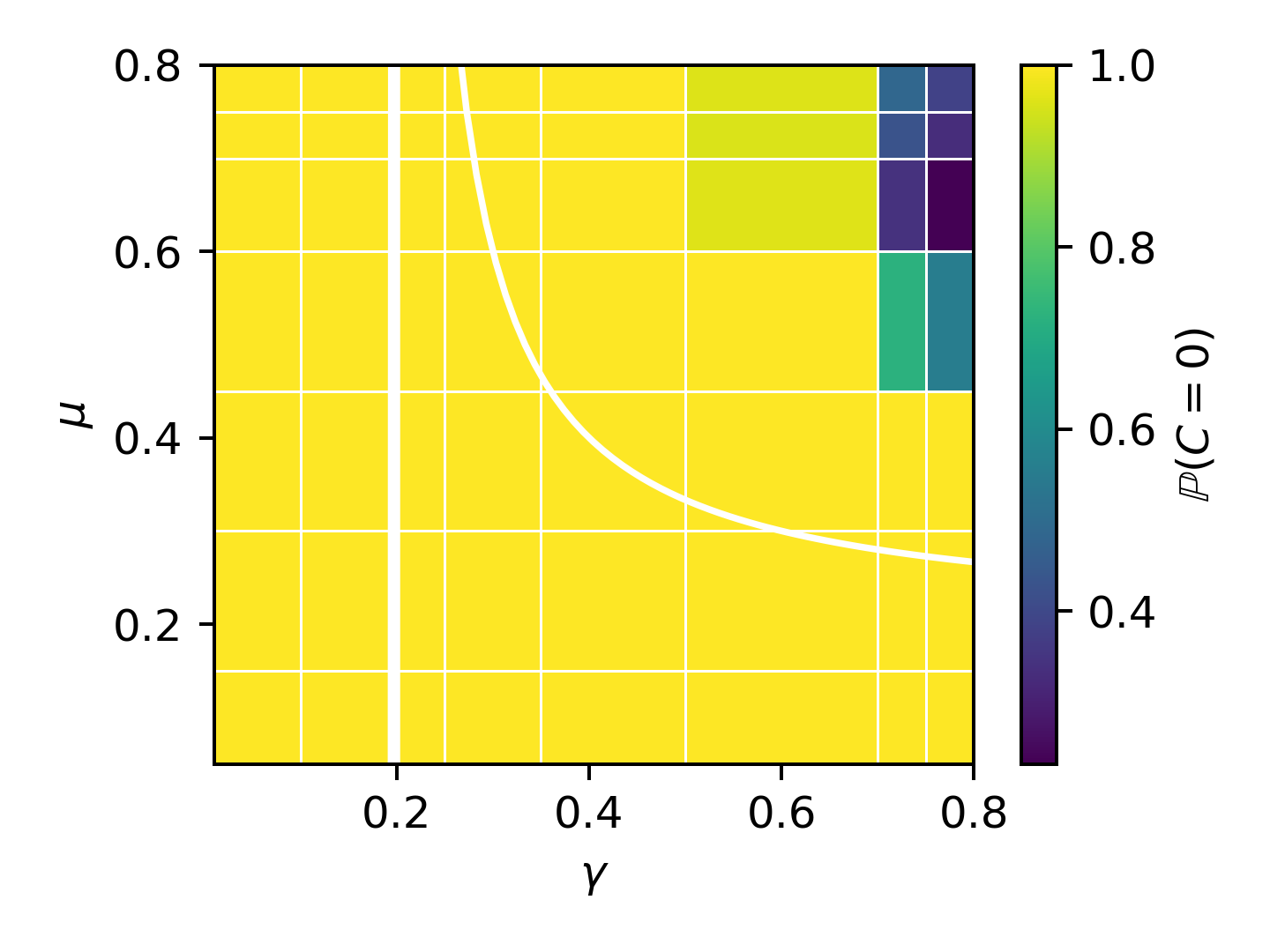}
\caption{$C$, for $N=100$}
    \end{subfigure}%
    \begin{subfigure}{0.45\textwidth}
         \centering           
\includegraphics[width = \textwidth]{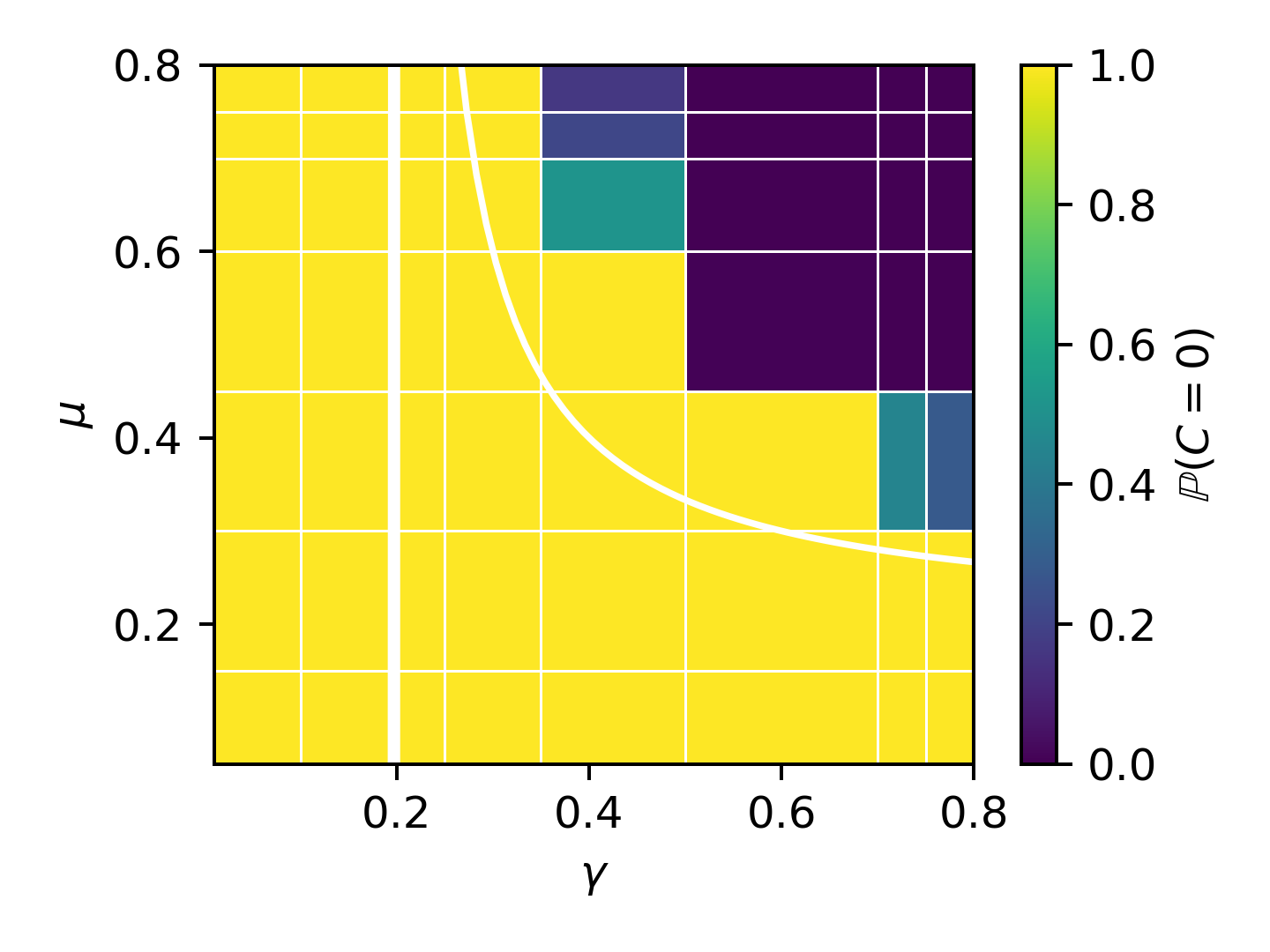}
\caption{$C$, for $N=1000$}
    \end{subfigure}%
    \\
    \begin{subfigure}{0.45\textwidth}
        \centering    
\includegraphics[width = \textwidth]{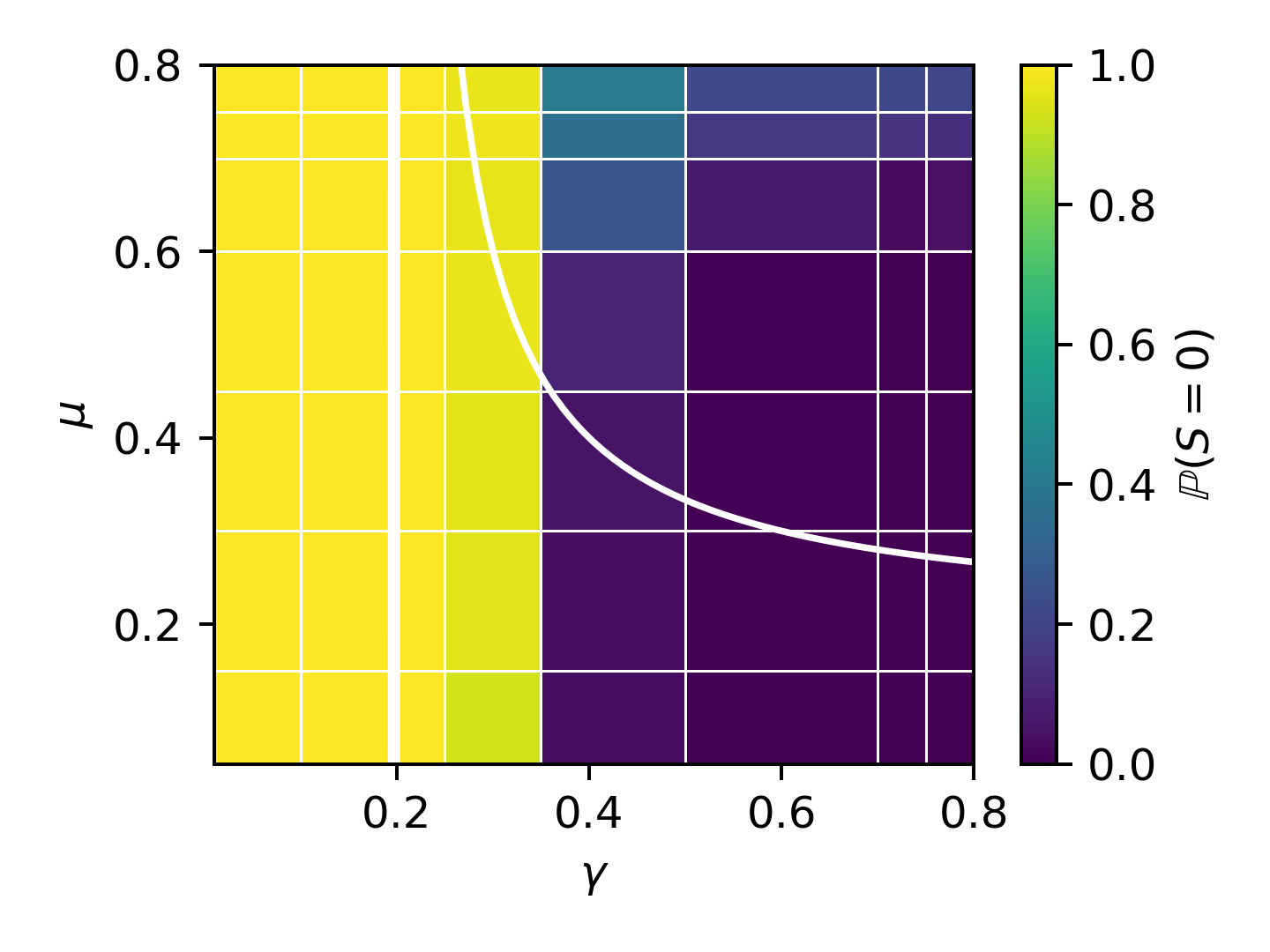}
\caption{$S$, for $N=100$}
    \end{subfigure}%
    \begin{subfigure}{0.45\textwidth}
       \centering     
\includegraphics[width = \textwidth]{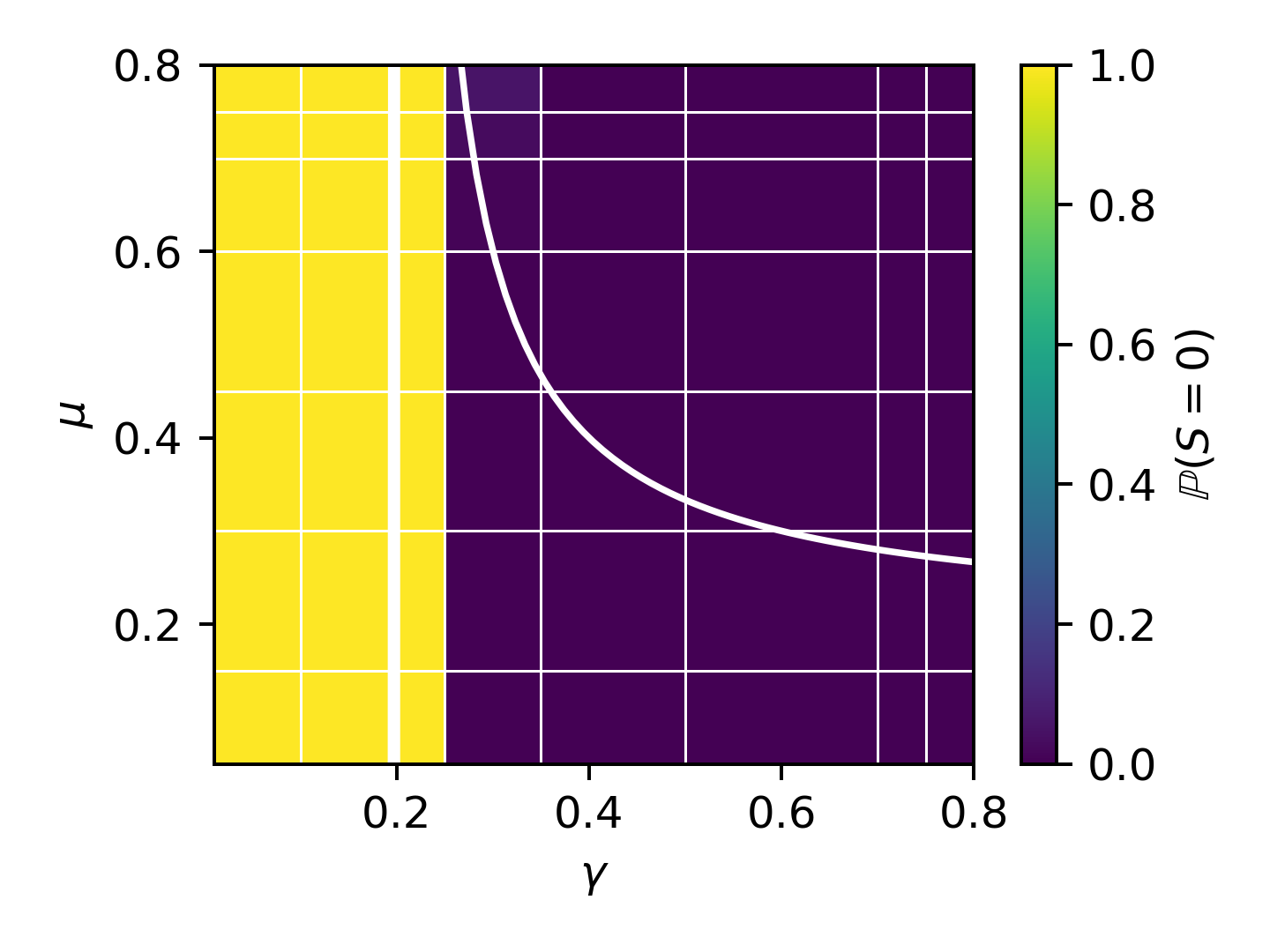}
\caption{$S$, for $N=1000$}
    \end{subfigure}%
    \caption{Extinction probabilities computed by simulation of the system with $\beta = 0.1$, $\nu=0.2$, for different values of $\mu$ and $\gamma$. The number of individuals in the population is also varied, as noted in the sub-caption. The white lines indicate the phase transition between the stability of stationary points.}
    \label{fig:Extinc_prob}
\end{figure}

In Figure~\ref{fig:Extinc_prob} we depict the computed extinction probability for the $C$ and $S$ compartments for the system with $\nu = 0.2,$ and $\beta = 0.1$. The remaining parameters were in the ranges: $\gamma\in\{0.01, 0.1, 0.25, 0.35, 0.5, 0.7, 0.75, 0.8\}$, and $\mu\in\{0.05,0.15, 0.3 ,0.45, 0.6, 0.7, 0.75, 0.8\}$. Each set of parameters was iterated 500 times. 

We see that as the population size increases, the dynamics behave more as predicted by their expectation (compartments not going extinct in regions where they are not expected to). This may be expected by the convergence of the diffusion approximation to the stochastic model as the population approaches infinity. Nonetheless, this indicates that for smaller populations (communities being modelled) are less susceptible to losing trust. Note that this pertains mainly to the situation in which only a small proportion of them would have lost trust at a time, rather than situations in which a large proportion of the small population loses trust. 

\section{Discussion}\label{sec:discussion}
In this final section we recapitulate the conclusions and insights gained from the analysis of our compartmental model. We also discuss the limitations of the setup chosen, as well as some directions for future work.

\subsection{Recapitulation of model and evaluation methodologies}
So as to better understand trust in society, we have developed a compartmental model that describes the process of individuals gaining or losing trust in institutions like governments. In our specific model, the population is split into three compartments: trusters ($T$), skeptics ($S$), and those having lost all trust, the doubters ($C$). By social influence, individuals may switch between compartments, but the switching mechanism includes a degree of bounded confidence: individuals in compartment $T$ are not influenced directly by individuals in $C$ or vice-versa. Thus, from $T$ one can jump to $S$, and from $S$ to both $T$ and $C$. This means that we exclude a transition from $C$ back to $S$, reflecting the adage that trust, once lost, is hard to regain.

In the literature,  the norm is to analyze such compartmental models in the fluid regime, which in practice means that one captures the system's mean behavior through a set of coupled differential equations. In our work we have the more ambitious objective to also shed light on the fluctuations around the fluid path. Being able to quantify the likelihood of such fluctuations, one can get a handle on \textit{e.g.}\ the probability of the compartment $S$ or $C$ becoming extinct, which is a quantity that one cannot analyze at the fluid limit level. 
We also stress that fluid limits provide a suitable tool when the population is large, and the evolution of the underlying compartmental model is strongly concentrated around the fluid path; in situations that the population considered is moderately sized, fluctuations matter, making the diffusion limit (or stochastic simulation) a more appropriate tool.

\subsection{Insights}

Firstly the model predicts that, if one could compare populations which differ \textit{only} in life-expectancy ($\nu$ and $\Lambda$), the population with a longer life-expectancy will have a greater proportion of individuals who have lost trust in their government. This is of course not straightforward to check on real world data because as life-expectancy increases in a country this may be related to the better operation of that country's government, and thus likely to influence other parameters of the model.
Though not sufficient evidence to conclude that this phenomenon is at play, we do see the trust in government in the US showing a downward trend~\cite{owid} since 1958, while the life expectancy has been steadily increasing since then. This insight is attainable from the fluid approximation.

Secondly, the model predicts that a larger population size is beneficial for doubters under parameter values for which $\bm{X}_3$ the mixed point is stable. In particular, as the population increases the probability of the doubter group dying out decreases and the expected proportion of the population in the doubter group increases. This happens because as the population size increases the stochastic system behaves more like the fluid limit approximation. Drawing this insight is not possible with only the fluid approximation but required our stochastic model. Importantly, we know \emph{a priori} that a larger population behaves more like its fluid approximation but we require the stochastic model in order to determine \emph{how} the pre-limit model differs from its expectation.

Thirdly, the model suggests that the relationship between the number doubters ($C$) and the coefficient of the rate at which individuals transition to this group ($\mu$) from the skeptics is not monotone. Thus from the perspective of a group of doubters (such as a conspiracy theorist group), there is an incentive not to be \textit{too} convincing in winning people over. This risks diminishing the skeptic population too quickly and being completely cut-off from the mainstream trusters if the skeptics die out. 

A final insight, which is not qualitative but rather methodological, is that studying the stochastic counterpart to the deterministic ODE models is not an overly complicated task. In particular the fluid and diffusion approximation to the stochastic model are useful in cases when the model predicts extinction of a compartment in expectation or when extinction is impossible (the techniques of fluid and diffusion approximation may be readily applied to the model studied by Wang~\cite{Wang2020} for instance). While for models which are susceptible to extinction of a compartment, Gillespie simulation is a useful tool which is relatively efficient and straightforward to implement.


\subsection{Directions for future research}

Like most belief dynamics models based on the analogy of epidemics~\cite{Olsson2024}, this model does not account for the effect of similarity between individuals on the likelihood of successful transmission of trust (or doubt). Based on the success of the epidemics analogy as well as its limitations, valuable future work might extended the model to include; an `immunized' state from which trust can never be lost, a `backlash' effect whereby the discussion between a skeptic and a doubter results in the skeptic transitioning back to the trust compartment, or indeed a sub-classification of compartments by which similarity between agents may be taken into account. These extensions fall outside the scope of the current paper\footnote{The scope of the current paper being a presentation and analysis of the simplest compartmental model for the dynamics of trust in society that includes bounded confidence. The inclusion of the extensions would add a host of variables whose effect would have to be disentangled from the baseline we present here.}, though the techniques applied to the current model are straightforward to implement to all of these extensions.

The model may be fitted to data should this be available. For instance the work of McCartney and Glass~\cite{McCartney2015} (on the number of church goers in Northern Ireland) may be extended with more data, and by the dependence of the stochastic model on the population size. This may give insights beyond the expected value, and to questions related to the probability of certain events of interest. For instance, there may be a level of participation required to sustain the activities of an institution such as the church; then this model could predict the likelihood of participation dropping below that level by a specified date.

The current model relies on the assumption that all transitions are persuasive in nature. It may be interesting to investigate the possibility of self-induced transitions which model the possibility of individuals losing or gaining trust in their government based on reasoning (such as modeled in \textit{e.g.}~\cite{Meylahn2024tat}) rather than social influence. This would bring the model closer to the one studied by Wang~\cite{Wang2020} on the spread of an irrational idea.
\bibliography{refs} 
\bibliographystyle{unsrt}
\appendix
\section{Verification of conditions for fluid and diffusion approximation}\label{ap:verification}
In this section we verify the conditions for the fluid and diffusion approximation stated in \S\ref{sec:methods}. In general this is straightforward because the model we set up closely follows the template in Kurtz~\cite{Kurtz1981}, though there are some issues that arise due to the introduction of a birth and death rate (making the population size variable and in fact unbounded).

\subsection{Fluid approximation conditions}\label{ap:verificationF}
The first condition (F.1) for the fluid limit to hold states that $\lim_N\to\infty \bm{X}_N(0) = \bm{x}_0$. This condition is met for our analysis by choosing a deterministic initial condition $\bm{X}_N(0)$ which implies $\bm{X}_N(0) = \bm{x}_0$ with probability~1.

The second condition (F.2) on the fluid limit approximation is $\sum_{\bm{\ell}\in\mathsf{L}}\sup_{\bm{x}\in\bm{K}}\beta_{\bm{\ell}}(\bm{x}) <\infty$, for all compact $\bm{K}\subset \mathbb{R}^d.$ The condition loosely translates to a requirement that the rates at which transitions occur be bounded for all possible states. Let $\bm{K}$ be a compact set in $\mathbb{R}^3$ (we remind the reader of the scaling by $N$ taking the process from $\mathbb{N}^3$ to $\mathbb{R}^3$). Then each transition $\beta_{\rate}$ attains its (not unique) maximum rate on this set on its boundary. Observe that each transition occurs at either a constant rate or proportional to $X_i$ or $X_iX_j$ for some $i\neq j\in \{1,2,3\}$. Thus the supremum is attained by maximizing the value of $X_i$ or $X_i$ and $X_j$ for the appropriate $i,j\in\{1,2,3\}$, and this supremum is finite. The finite sum (there are 7 transitions) of finite numbers is finite and so this condition is met. 

The final condition (F.3) of the fluid limit $\exists M_K>0: |\bm{F}(\bm{x}_1)-\bm{F}(\bm{x}_2)|\leq M_K|\bm{x}_1-\bm{x}_2|$, $\bm{x}_1,\bm{x}_2\in \bm{K}$, for all compact $\bm{K}\subset\mathbb{R}^3$ implies continuity of sorts. 
The difference on the left hand side of the inequality can be written as follows,
\begin{equation}
    |\bm{F}(\bm{x}_1)-\bm{F}(\bm{x}_2)| = 
  \Bigg\vert  \begin{bmatrix}
        \nu (T_2-T_1) +(\alpha-\beta)(T_2S_2-T_1S_1)\\
        -(\alpha-\beta)(T_2S_2-T_1S_1) +\nu (S_2-S_1) +\mu (S_2C_2-S_1C_1)\\
        -\mu (S_2C_2-S_1C_1) +\nu (C_2-C_1)
    \end{bmatrix}\Bigg\vert.
\end{equation}
Meanwhile the right hand side is
\begin{equation}
   M_K|\bm{x}_1-\bm{x}_2|=M_K \sqrt{(T_1-T_2)^2+(S_1-S_2)^2+(C_1-C_2)^2}.
\end{equation}
In the case of $\bm{x}_1=\bm{x}_2$, the requirement is met for any $M_K$ because then both the left and right hand side of the condition are zero. We note that for specific $\bm{x}_1\neq\bm{x}_2\in\bm{K}$ we have equality $|\bm{F}(\bm{x}_1)-\bm{F}(\bm{x}_2)| = M_K|\bm{x}_1-\bm{x}_2|$ when 
\begin{equation}\label{eq:MK}
    M_K (\bm{x}_1,\bm{x}_2) =  \frac{|\bm{F}(\bm{x}_1)-\bm{F}(\bm{x}_2)|}{\sqrt{(T_1-T_2)^2+(S_1-S_2)^2+(C_1-C_2)^2}}.
\end{equation}
Division is allowed because $\bm{x}_1\neq \bm{x}_2$ and so the denominator is positive. So for any compact $\bm{K}\in\mathbb{R}^3$ we simply choose the $M_K$ which maximizes (\ref{eq:MK}) as long as we know that $|\bm{F}(\bm{x}_1)-\bm{F}(\bm{x}_2)|<\infty$. 


\subsection{Diffusion approximation conditions}\label{ap:verificationD}
Similarly to (F.1), the first condition (D.1) on the diffusion limit is verified immediately by starting the system from a deterministic state $\bm{X}_N(0)$. Thus $\lim_{N\to\infty}\sqrt{N}|\bm{X}_N(0)-x_0|=0$ almost surely as required.

The second condition (D.2) on the diffusion approximation is similar to the second condition of the fluid approximation with the addition of a multiplication of each term by $|\rate|^2$: $\sum_{\bm{\ell}\in\mathsf{L}}|\bm{\ell}|^2\sup_{\bm{x}\in\bm{K}}\beta_{\bm{\ell}}(\bm{x}) <\infty$, for all compact $\bm{K}\subset\mathbb{R}^d$. The transitions in our model are such that $|\rate|^2$ is at most $2$ because either only one compartment increases/decreases by one (by an arrival or an exit), or one compartment increases by one while another decreases by one (by a transition). Thus the terms in the finite sum are still finite, and the condition is verified.

The condition D.3 is technically {\it not} met: there are the entries consisting of terms proportional to $x_1x_2$ and $x_2x_3$ in ${\bm F}({\bm x})$, such that the matrix $\nabla \bm{F}(\bm{x})$ contains entries with terms proportional to $x_1$, $x_2$, and $x_3$,
which are continuous, but not bounded. This problem can be circumvented by truncating the state space at an appropriately high level, as follows.
\begin{itemize}
\item 
Let $\hat{\bm X}_{N,\kappa}(\cdot)$ have the same generator as $\hat{\bm X}_{N}(\cdot)$, but with the sum of the three components of ${\bm X}(\cdot)$ truncated at level $N\kappa$, with $\kappa$ larger than the sum of the entries of ${\bm x}_0$ (below denoted by $z$). This process $\hat{\bm X}_{N,\kappa}(\cdot)$ clearly satisfies the condition D.3, as any continuous function on a bounded domain is bounded. Define by $\hat{\bm X}_\kappa(\cdot)$ the counterpart of $\hat{\bm X}(\cdot)$, as defined through \eqref{eq:dif_sde}, in this truncated model.
\item 
Let, for an arbitrarily chosen $T>0$,  ${\mathscr E}_N$ be the event that the total network population remains below $N\kappa$ for all $t\in[0,T]$, so that
\begin{equation}
    {\mathfrak p}_{N,\kappa}:={\mathbb P}({\mathscr E}_N^{\rm c})= {\mathbb P}(\exists t\in [0,T]:T(t)+S(t)+C(t) \geqslant N\kappa).   
\end{equation}
In order to make sure that ${\mathscr E}_N^{\rm c}$ applies, a necessary condition is that, with $P(\cdot)$ a Poisson process with rate $N\Lambda$, that $P(T) \geqslant N(\kappa-z)$. Hence, applying the Chernoff bound, for any $\kappa\geqslant z+\Lambda T$,
\begin{equation}\label{eq:pnk}
 {\mathfrak p}_{N,\kappa} = \sum_{i=N\kappa}^\infty e^{-N\Lambda T} \frac{(N\Lambda T)^i}{i!}\leqslant e^{-NI(\kappa-z)},    
\end{equation}
with $I(\cdot)$ the Legendre transform $I(a):= a\log(a)-a\log(\Lambda T)-a+\Lambda T$. 
\item
It thus follows that, for any given set $A$ of paths on $[0,T]$,
\begin{align}
    {\mathbb P}\left((\hat{\bm X}_N(t))_{t\in[0,T]}\in A, {\mathscr E}_N\right)&\leqslant  {\mathbb P}\left((\hat{\bm X}_N(t))_{t\in[0,T]}\in A\right)\\
    &\leqslant  {\mathbb P}\left((\hat{\bm X}_N(t))_{t\in[0,T]}\in A, {\mathscr E}_N\right) + {\mathfrak p}_{N,\kappa},
\end{align}
so that, by \eqref{eq:pnk},
\begin{equation}
\lim_{N\to\infty}{\mathbb P}\left((\hat{\bm X}_N(t))_{t\in[0,T]}\in A, {\mathscr E}_N\right) =     \lim_{N\to\infty} {\mathbb P}\left((\hat{\bm X}_N(t))_{t\in[0,T]}\in A\right).
\end{equation}
\item
Now choose $\kappa$ sufficiently large so that the full fluid path is such that, given it starts at ${\bm x}_0$, the sum of the entries stays below $\kappa$ during $[0,T]$. This means that 
\begin{align*}
  \lim_{N\to\infty}{\mathbb P}\left((\hat{\bm X}_N(t))_{t\in[0,T]}\in A, {\mathscr E}_N\right) &=    
  \lim_{N\to\infty}{\mathbb P}\left((\hat{\bm X}_{N,\kappa}(t))_{t\in[0,T]}\in A\right)\\& = {\mathbb P}\left((\hat{\bm X}_\kappa(t))_{t\in[0,T]}\in A\right)= {\mathbb P}\left((\hat{\bm X}(t))_{t\in[0,T]}\in A\right);
\end{align*}
the first equality is because on ${\mathscr E}_N$ the processes $\hat{\bm X}_N(t))_{t\in[0,T]}$ and $\hat{\bm X}_{N,\kappa}(t))_{t\in[0,T]}$ coincide, the second equality because condition D.3 applies to $\hat{\bm X}_{N,\kappa}(\cdot)$, and the third equality because $\hat{\bm X}_N(\cdot)$ and $\hat{\bm X}_{N,\kappa}(\cdot)$ have the same fluid paths. 
\end{itemize}
Upon combining the above, we conclude, for any given set $A$ of paths on $[0,T]$,
\begin{equation}
     \lim_{N\to\infty} {\mathbb P}\left((\hat{\bm X}_N(t))_{t\in[0,T]}\in A\right) = {\mathbb P}\left((\hat{\bm X}(t))_{t\in[0,T]}\in A\right),
\end{equation}
as desired.
\section{Stability analysis}\label{app:ODE_stab}
The nonlinear system may be approximated close to the fixed points by the linearized system:
\begin{equation}
    \dot{\bm{X}} = A \bm{X}(t),
\end{equation}
where $A$ is the Jacobian:
\begin{equation}
    A:= \begin{bmatrix}
        \frac{d\bm{X}_1}{dT}, \frac{d\bm{X}_1}{dS}\\
        \frac{d\bm{X}_2}{dT}, \frac{d\bm{X}_2}{dS}
    \end{bmatrix}.
\end{equation}
In particular we get the matrices
\begin{equation}
    A_i= \begin{bmatrix}
        -\nu -\gamma S_i, & -\gamma T_i\\
        (\gamma +\mu)S_i, &(\gamma+\mu)T_i -(\nu+\mu)+\mu S_i
    \end{bmatrix},
\end{equation}
corresponding to the fixed points $i=1,2,3$. For each fixed point we proceed to compute the trace ($\tau_i$) and the determinant ($\Delta_i$) (tabulated in Table~\ref{tab:stab}). Following the procedure meticulously explained in the book by Strogatz~\cite{Strogatz2024} we can determine the stability and type of the fixed points by checking whether $\tau_i<0$ (stable if true) and $\tau_i^2-4\Delta_i>0$ (node if true). 

Intuitively speaking, a stable node is an attracting fixed point to which the solution of the dynamic system converges to as $t\to\infty$. The path toward the node may be straight or curved but is without oscillation (see for example Figures~\ref{fig:x1}--\ref{fig:x3node}). A stable spiral on the other hand involves damped oscillation during the convergence toward the attracting fixed point at the center of the oscillations (see Figure~\ref{fig:x3spiral}).

Note that on the boundaries between the regions where two fixed points are stable (as depicted in Figure~\ref{fig:Stability}), the fixed points on either side of the boundary are equal. That is when $\nu=\gamma$, then $\bm{X}_1=\bm{X}_2$. Similarly, when $\mu=\nu\gamma / (\gamma-\nu)$, $\bm{X}_3=\bm{X}_2$.

\begin{table}[hb]
    \centering
    \begin{tabular}{|l|c|c|c|}
    \hline
   $i$ & $\bm{X}_i$ & $\tau_i$ & $\Delta_i$\\
    \hline
     1 &  $(1,0,0)$ & $\gamma -2\nu$ & $\nu^2 - \nu\gamma$ \\
      2 & $(\nu/\gamma,1-\nu/\gamma,0)$  & $-\gamma+\mu-\mu\nu/\gamma$ & $2\nu\mu +\nu\gamma-\mu\gamma-\nu^2(1+\mu/\gamma)$\\
      3 & $(\frac{\mu}{\gamma+\mu},\frac{\nu}{\mu},\frac{\gamma }{\gamma+\mu}-\frac{\nu}{\mu} )$   & $-\gamma\nu/\mu $ & $\gamma\nu-\nu^2-\gamma\nu^2/\mu$\\
      \hline
    \end{tabular}
    \caption{Trace and determinant at fixed points.}
    \label{tab:stab}
\end{table}
\subsection{Fixed point 1}
The first fixed point $\bm{X}_1=(1,0,0)$ is a stable node when $\gamma\leq\nu.$ To see this first note that $\tau_1$ is negative for all $\gamma<2\nu$ which is also true when $\gamma<\nu$. Secondly, $\Delta_1$ is negative whenever $\nu<\gamma$. When $\Delta_i<0$ this implies that fixed point $i$ is a saddle point, while $\tau_i<0$ implies stability. Thus for $\bm{X}_i$ to not be a saddle point, and stable, we require $\tau_1<0$ and $\Delta_1>0$. This is true for $\nu<\gamma$. To see that $\bm{X}_1$ is a node we inspect $\tau_1^2-4\Delta_1:$
\begin{align*}
    \tau_1^2-4\Delta_1 &= (\gamma - 2\nu)^2 - 4(\nu^2 - \nu\gamma)=\gamma^2.
\end{align*}
This is positive whenever $\gamma>0$ which is true by assumption. Thus $\bm{X}_1$ is a node rather than a spiral (which would have been the case if $\tau_1^2-4\Delta_1<0$).

\subsection{Fixed point 2}
The second fixed point $\bm{X}_2= (\nu/\gamma,1-\nu/\gamma,0)$ is a saddle point when $\gamma<\nu$ and when $\mu>\nu\gamma/(\gamma-\nu)$. On the contrary it is a stable node when $\gamma>\nu$ and $\mu<\nu\gamma/(\gamma-\nu)$.

By the assumption that the entry rate and the exit rate of the system are equal, and that we start with a total population of 1, we immediately see that $\nu>\gamma$ would imply that $T>1$ and $S<0$ which is not possible. Manipulating the expression for $\tau_2<0$ we see that this is true when:
\begin{equation*}
    \mu(1-\nu/\gamma)<\gamma,
\end{equation*}
which requires,
\begin{equation*}
    \mu <\gamma/(1-\nu/\gamma).
\end{equation*}
This last expression can be simplified to $\mu<\gamma^2/(\gamma-\nu)$. However, we also see that $\bm{X}_2$ is a saddle point whenever $\Delta_2<0$. This is the case for 
\begin{equation*}
    \mu > \frac{-\nu^2+\nu\gamma}{\gamma-2\nu+\nu^2/\gamma}.
\end{equation*}
Taking out a factor of $\nu$ on the right hand side (numerator and denominator) we get
\begin{equation*}
     \mu > \frac{\gamma-\nu}{\gamma/\nu-2+\nu/\gamma}.
\end{equation*}
Finally this can be rewritten to state that $\bm{X}_2$ is a saddle point (because $\Delta_2<0$) when $\mu>\nu\gamma/(\gamma-\nu)$. Conversely $\Delta_2>0$ when $\mu<\nu\gamma/(\gamma-\nu)$, and because under this condition $\tau_2<0$ (because $\nu<\gamma$ by assumption) $\bm{X}_2$ is a stable fixed point. It is straightforward to check that $\tau_2^2-4\Delta_2$ is positive under these conditions, meaning that the type of this fixed point is a node.

\subsection{Fixed point 3}
Fixed point $\bm{X}_3= ((\frac{\mu}{\gamma+\mu},\frac{\nu}{\mu},\frac{\gamma }{\gamma+\mu}-\frac{\nu}{\mu} ))$ is stable when $\mu>\nu\gamma/(\gamma-\nu)$. To see that this is true observe that $\tau_3=\gamma\nu/\mu<0$ for all parameter values considered. This means that as soon as $\Delta_3>0$, $\bm{X}_3$ switches from a saddle point to a stable fixed point. Thus $\bm{X}_3$ is stable when:
\begin{equation}
    \gamma\nu - \nu^2-\gamma\nu^2/\mu>0,
\end{equation}
dividing all terms by $\nu$ does not change the sign of either side because $\nu>0$, thus we can rewrite this as:
\begin{equation}
    \mu(\gamma-\nu)-\gamma\nu>0.
\end{equation}
This in turn can be rearranged to get the stated inequality: $\mu>\gamma\nu/(\gamma-\nu)$.

To determine the type of fixed point we check the sign of $\tau_3^2-4\Delta_3$. If it is positive (negative) then $\bm{X}_3$ is a node (spiral). $\tau_3^2-4\Delta_3<0:$
\begin{equation}
    \frac{\gamma^2\nu^2}{\mu^2}-4\gamma\nu + 4\nu^2+\frac{4\gamma\nu^2}{\mu}<0,
\end{equation}
multiplying through by $\mu^2/\nu$ yields:
\begin{equation}
    \gamma^2\nu-4\gamma\mu^2+4\nu\mu^2+4\mu\gamma\nu<0.
\end{equation}
By rearranging terms we notice that this is a quadratic polynomial in $\mu$:
\begin{equation}
    \mu^2(4\nu-4\gamma) + \mu (4\gamma\nu)+\gamma^2\nu<0,
\end{equation}
which can readily be solved for roots using the quadratic formula:
\begin{equation}
    \mu_1 =\frac{\gamma  \nu -\gamma ^{3/2} \sqrt{\nu }}{2 (\gamma -\nu )}, \quad \mu_2 =\frac{\gamma\nu+\gamma ^{3/2} \sqrt{\nu }}{2 (\gamma -\nu )} .
\end{equation}
Thus the condition under which $\bm{X}_3$ is a spiral is
\begin{equation}
    (\mu-\mu_1)(\mu-\mu_2)<0.
\end{equation}
In the regime of interest where $\gamma>\nu$, it can be checked that $\mu_1$ is always negative. Thus the sign of the entire expression is determined by $\mu-\mu_2$. Because $\mu_2$ is always positive (again by $\gamma>\nu$) we get that $\bm{X}_3$ is a spiral when $\mu>\mu_2$, a node when $\mu<\mu_2$ and a star when $\mu=\mu_2$.

\end{document}